\documentclass[11pt,a4paper]{article}
\usepackage{epsfig}
\usepackage[T1]{fontenc}    
\usepackage{graphics}
\usepackage{graphicx}
\usepackage{pstricks,pst-coil,pst-fill,pst-plot}
\usepackage[fleqn]{amsmath}    
\usepackage{amssymb}    
\usepackage{amsfonts}   
\usepackage{verbatim}   
\usepackage{mathrsfs}   
\usepackage{dsfont}
\usepackage{euscript}
\usepackage{yfonts}
\usepackage{enumerate}     
\usepackage{txfonts}
\usepackage{marvosym}
\usepackage{vmargin}        
\usepackage{wasysym}		

\setmarginsrb{1.8cm}{2cm}{1.8cm}{2cm}{1cm}{1cm}{1cm}{1.6cm}
 \makeatletter
 \@addtoreset{equation}{section}
 \makeatother

\providecommand{\bysame}{\leavevmode\hbox to3em{\hrulefill}\thinspace}
\providecommand{\MR}{\relax\ifhmode\unskip\space\fi MR }

\providecommand{\href}[2]{#2}

\let\ua=\uparrow
\let\da=\downarrow
\let\tend=\rightarrow

\long\def\symbolfootnote[#1]#2{\begingroup%
\def\thefootnote{\fnsymbol{footnote}}\footnote[#1]{#2}\endgroup}

\newtheorem{theorem}{Theorem}[section]
\newtheorem{prop}{Proposition}[section]

\newtheorem{defin}{Definition}[section]

\newtheorem{lemme}{Lemma}[section]

\def\Proof{\medskip\noindent {\it Proof --- \ }}

\def\qed{\hfill\rule{2mm}{2mm}}

\newcommand\beq{\begin{equation}}
\newcommand\enq{\end{equation}}
\newcommand\bem{\begin{multline}}
\newcommand\enm{\end{multline}}

\def\beqa{\begin{eqnarray}}
\def\eeqa{\end{eqnarray}}
\def\ba{\begin{array}}
\def\ea{\end{array}}
\def\det{\operatorname{det}}

\newcommand{\f}[2]{{\ensuremath{%
    \mathchoice%
    {\dfrac{#1}{#2}}
    {\dfrac{#1}{#2}}
    {\frac{#1}{#2}}
    {\frac{#1}{#2}}
}}}
\newcommand{\tf}[2]{\ensuremath{#1/#2}}
\newcommand{\pa}[1]{\ensuremath{\left(#1\right)}}

\def\a{\alpha}

\def\be{\beta}
\def\ga{\gamma}
\def\Ga{\Gamma}

\def\de{\delta}

\def\eps{\epsilon}
\def\veps{\varepsilon}
\def\la{\lambda}

\def\sg{\sigma}
\def\vsg{\varsigma}
\def\Sg{\Sigma}

\def\Ups{\Upsilon}

\newcommand{\mc}[1]{\ensuremath{\mathcal{#1}}}
\newcommand{\mf}[1]{\ensuremath{\mathfrak{#1}}}
\newcommand{\msc}[1]{\ensuremath{\mathscr{#1}}}

\newcommand{\bs}[1]{\ensuremath{\boldsymbol{#1}}}

\DeclareFontFamily{OT1}{pzc}{}
\DeclareFontShape{OT1}{pzc}{m}{it}{<-> s * [1.10] pzcmi7t}{}
\DeclareMathAlphabet{\mathpzc}{OT1}{pzc}{m}{it}

\def \i{ \mathrm i}

\newcommand{\ov}[1]{\ensuremath{\overline{#1}}}
\newcommand{\wt}[1]{\ensuremath{\widetilde{#1}}}
\newcommand{\wh}[1]{\ensuremath{\widehat{#1}}}

\newcommand{\Int}[2]{\ensuremath{\int\limits_{#1}^{#2}}}
\newcommand{\Oint}[2]{\ensuremath{\oint\limits_{#1}^{#2}}}

\newcommand{\sul}[2]{\ensuremath{\sum\limits_{#1}^{#2}}}

\newcommand{\R}{\ensuremath{\mathbb{R}}}
\newcommand{\Cx}{\ensuremath{\mathbb{C}}}

\newcommand{\Dp}[1]{\ensuremath{\partial_{#1}}}

\newcommand{\limit}[2]{\ensuremath{\underset{#1 \tend #2}{\longrightarrow} }}

\newcommand{\ex}[1]{\ensuremath{\e{e}^{#1}}}

\newcommand{\abs}[1]{\ensuremath{\left| #1 \right|}}

\newcommand{\norm}[1]{\ensuremath{|| #1 ||}}

\newcommand{\dd}{\mathrm{d}}
\newcommand{\e}[1]{\ensuremath{\mathrm{#1}}}

\newcommand{\intff}[2]{\ensuremath{ [  #1 \,; #2 ] }}

\newcommand{\intof}[2]{\ensuremath{ ]  #1 \,; #2 ] }}
\newcommand{\intoo}[2]{\ensuremath{ ]  #1 \,; #2 [ }}

\newcommand{\intn}[2]{\ensuremath{[\![ \, #1 \,;\, #2 \,]\!]}}

\begin{document}




\begin{center}
\begin{LARGE}
{\bf Large-$x$ analysis of an operator valued Riemann--Hilbert problem}
\end{LARGE}

\vspace{7mm}

{\bf A.~R.~Its}\symbolfootnote[1]{Indiana University Purdue University Indianapolis, Department of Mathematics, Indianapolis, USA, itsa@math.iupui.edu} , 
{\bf K.~K.~Kozlowski}\symbolfootnote[2]{Universit\'{e} de Bourgogne, Institut de Math\'{e}matiques de Bourgogne, UMR 5584 du CNRS, France,
karol.kozlowski@u-bourgogne.fr}. 
\par

\vspace{40pt}

\centerline{\bf Abstract} \vspace{1cm}
The purpose of this paper is to push forward the theory of operator-valued Riemann Hilbert problems and demonstrate their 
effectiveness in respect to the implementation of a non-linear steepest descent method \textit{\'{a} la} Deift-Zhou. 
In the present paper,  we demonstrate that the operator-valued Riemann--Hilbert problem arising in the characterisation
of so-called $c$-shifted integrable integral operators allows one to extract the large-$x$ asymptotics of the Fredholm determinant 
associated with such operators. 

\parbox{12cm}{\small}

\end{center}

\vspace{40pt}

\section{Introduction}

The term integrable integral operator refers to a specific class of integral operators $I+V$ whose integral kernel 
takes the form 
\beq
V(\la,\mu) \; = \; \f{ \sum_{a=1}^{N} e_a(\la)f_a(\mu)  }{ \la - \mu } \qquad \e{with} \qquad \sul{a=1}{N} e_a(\la)f_a(\la) \; = \; 0
\label{ecriture noyau op int de rg fini}
\enq
where $e_a$ , $f_a$, $a=1,\dots,N$ are functions whose regularity depends on the functional space on which 
the operator acts. The quite specific structure of their kernels endows integrable integral operators with numerous 
properties allowing one, in particular, for the construction of the resolvent kernel or computation of the Fredholm determinant 
of $I+V$ in terms of a solution to a specific $N\times N$ matrix valued Riemann--Hilbert problem \cite{ItsIzerginKorepinSlavnovDifferentialeqnsforCorrelationfunctions}. 
We remind that the jump matrix for this Riemann--Hilbert problem is built out of the functions $e_a$ and $f_a$, $a=1,\dots,N$. 

Despite the specific form \eqref{ecriture noyau op int de rg fini} imposed on the kernel of integrable integral operators, such 
operators still arise in many concrete problems of  mathematical physics. 
The Fredholm determinants of specific instances of such operators describe numerous observables, be it in random matrix theory
-gap probabilities in the bulk or edge of the spectrum \cite{DeiftGioevProodUniversalityEdgeBeta124RHPApproach,DeiftKriechMcLaughVenakZhouOrthogonalPlyExponWeights,GaudinGUELevelSpacingasSineKernel}
- or quantum integrable models -correlation functions of products of local operators \cite{KorepinSlavnovTimeDepCorrImpBoseGas,SlavnovPDE4MultiPtsFreeNLSM}- to name a few.

One can, in fact, consider more general integrable integral operators than those described by \eqref{ecriture noyau op int de rg fini}. 
To generalise the formula, it is enough to replace the discreet variable $a\in \{1,\dots,N\}$ labelling the functions $e_a$ and $f_a$
by a variable $s$ living in some measure space ($X,\nu$). One then replaces the discreet and finite sum in  \eqref{ecriture noyau op int de rg fini}
by an integral \textit{versus} $ \dd \nu$:
\beq
V(\la,\mu) \; = \; \f{ \int_{X}{} e(\la;s)f(\mu;s) \cdot \dd \nu(s) }{ \la - \mu } \qquad \e{with} \qquad  \int_{X}{} e(\la;s)f(\la;s) \cdot \dd \nu(s)   \; = \; 0 \;. 
\label{introduction class op val int int ops}
\enq

Particular, examples of such more general integrable integral operators arose in the context 
of studying quantum integrable systems at generic value of their interaction strength \cite{FrahmItsKorepinRHPAnalysisforXXXbyDualFields,ItsIzerginKorepinSlavnovDifferentialeqnsforCorrelationfunctions, KojimaKorepinSlavnovNLSEDeterminatFormFactorAndDualFieldTempeAndTime}, \textit{viz}. 
away from their free fermion point. 
Independently from their existing applications, such more general integrable integral operators are of interest in their own right
precisely because of the much larger freedom in the form taken by their kernels and yet the possibility to study them by means
of Riemann--Hilbert problems. The price to pay, however, is the complication of the Riemann--Hilbert problem 
in that one no longer deals with a matrix valued one but rather an operator valued one. Still, in the early days of 
exploring the correlation functions in quantum integrable systems out of their free fermion point, certain properties of
Fredholm determinants of such more general operators were investigated on the basis of operator valued Riemann--Hilbert problems
which are associated with these kernels. The Riemann--Hilbert machinery allowed to construct systems of partial differential equations satisfied by specific
instances of such operators \cite{ItsIzerginKorepinSlavnovDifferentialeqnsforCorrelationfunctions,KojimaKorepinSlavnovNLSESystemPartialDiffEqnsDualFieldTempeAndTime,
KorepinSlavnovTempeNLSELongDistCompMoreTermsOldSeries,KorepinSlavnovRHPForNLSMOpertorValuedSettingInto}. It is also important to mention the work \cite{ItsSlavnovNLSTimeAndSpaceCorrDualFields}
where a formal non-linear steepest descent analysis of an oscillatory operator valued Riemann--Hilbert problem was carried out. 
This allowed the authors to extract the leading asymptotic behaviour in the large parameter out of the logarithm of the Fredholm determinant
at stake. However,  numerous technical difficulties (the operator nature of the scalar Riemann--Hilbert problem which 
arises in the very the first step of the analysis, construction of parametrices in terms of special functions with operator index,...) 
which could not have been overcome stopped, for almost 15 years, any activity related to an asymptotic analysis of operator valued Riemann--Hilbert problems.   

Recently in \cite{KozItsAsymptofcShiftedFredholms} we have proposed  a scheme allowing one to extract the large-$x$ asymptotic behaviour of the Fredholm determinant 
of so-called $c$-shifted integrable integral operators which belong to the class \eqref{introduction class op val int int ops}, with $X=\R^+\times \intn{1}{N}$
and $e_a, f_a$ depending on $x$ in an oscillatory way. 
The method of analysis we developed was completely disconnected from any use of the operator valued Riemann--Hilbert problem 
that is underlying to such $c$-shifted operators. Notwithstanding, the very fact that the large-$x$ behaviour of these determinants could have been extracted 
constituted a strong indication that there must exist a way for overcoming the technical difficulties that constituted a obstruction 
to the asymptotic analysis of operator valued Riemann--Hilbert problems. 

As a matter of fact, the recent progress in the field of Riemann--Hilbert problems brings new ideas and tools which allow one for an effective
asymptotic analysis of operator valued Riemann--Hilbert problems. The present paper is precisely devoted to demonstrating this fact. 
More precisely, we reformulate the original statement of an operator valued Riemann--Hilbert problem \cite{KorepinSlavnovRHPForNLSMOpertorValuedSettingInto}
what permits us to develop
a framework allowing one to discuss the solvability and uniqueness of solutions to operator valued Riemann--Hilbert problems. 
We demonstrate the effectiveness of our scheme by carrying out the large-parameter non-linear steepest descent analysis of an 
oscillatory operator-valued Riemann--Hilbert problem which can be though of as the operator-valued generalisation of the Riemann--Hilbert problem associated with 
the so-called generalised sine kernel \cite{KozKitMailSlaTerRHPapproachtoSuperSineKernel}.
Our analysis allows us to reproduce the results of \cite{KozItsAsymptofcShiftedFredholms} directly within the operator valued Riemann--Hilbert problem setting. 
We do stress that the main achievements of this paper is to overcome two technical difficulties which arose previously in the analysis 
of operator-valued Riemann--Hilbert problems:
\begin{itemize}
 
 \item primo,  we reduce the problem of constructing  solutions to  operator valued scalar Riemann--Hilbert problem with jump on $I$ to the one of inverting 
an integral operator acting on $L^{2}\big( \Ga(I), \dd z \big)$, in which $\Ga$ is a small counterclockwise loop around 
$I$.  

 \item Secundo, we strongly simplify the construction of local parametrices. More precisely, the setting we propose allows us to construct parameterices in terms of 
 special function (confluent hypergeometric functions in our case) whose auxiliary parameters are scalar-valued holomorphic functions and not 
 holomorphic functions taking values in some infinite dimensional Banach spaces,  as it was the case in \cite{ItsSlavnovNLSTimeAndSpaceCorrDualFields}.
 
\end{itemize}

In the present paper, we shall develop the formalism on the example of the below integrable integral operator on $L^2\big( \intff{a}{b} \big)$
of $c$-shifted type whose integral kernel reads
\beq
V(\la,\mu)  \; = \; 
\f{ \i c F(\la)  }{ 2\i \pi (\la-\mu) } \cdot \bigg\{ \f{ \ex{ \f{\i x}{2}[p(\la)-p(\mu)] } }{ (\la-\mu) + \i c } 
\, + \, \f{ \ex{ \f{\i x}{2}[p(\mu)-p(\la)] } }{ (\la-\mu) - \i c } \bigg\} \;. 
\label{definition noyau integrable V}
\enq
Throughout the paper, we shall assume that 
\begin{itemize}
 \item   $p(\intff{a}{b})\subset \R$ and that $p$ is a biholomorphism from an open neighbourhood $U$ 
 of $\intff{a}{b}$ in $\Cx$ onto some open neighbourhood of $\intff{p(a)}{p(b)}$ in $\Cx$ which furthermore satisfies $p^{\prime}_{\mid \intff{a}{b}}>0$ ;
\item $F$ is holomorphic on $U$ and satisfies $\big| \e{arg}\big(1+F(\la) \big) \big|<\pi$ for any $\la \in U$ . 
\end{itemize}

\noindent Our analysis allows us to prove the 
\begin{theorem}
\label{Theorem principal papier}
Let $p$ and $F$ be as described above and $V_0$ denote the integral operator on $L^{2}\big( \intff{a}{b} \big)$ whose integral kernl reads 
\beq
V_0(\la,\mu)  \; = \; \f{  F(\la)  }{  \pi (\la-\mu) } \cdot  \sin \Big( \f{ x}{2}[p(\la)-p(\mu)]  \Big) \;. 
\label{definition noyau integrable GSK}
\enq
Then the below ratio of Fredholm determinants admits the large-$x$ asymptotic behaviour
\beq
\f{ \det\big[ \e{id}\, + \, V \big] }{ \det\big[ \e{id}\, + \, V_0 \big]  } \; = \; 
\det_{ \Ga (\intff{a}{b} ) } \big[I + \mc{U}_+ \big] \cdot \det_{ \Ga(\intff{a}{b}) } \big[I+ \mc{U}_- \big] \cdot 
\Big( 1\,+\,\e{o}(1) \Big)
\enq
where $\mc{U}_{\pm}$ are integral operators on $L^{2}\big( \Ga(\intff{a}{b}) \big)$, with $\Ga$ being a small counterclockwise loop 
around the interval $\intff{a}{b}$. The integral kernels $U_{\pm}$ of $\mc{U}_{\pm}$ read
\beq
U_{\pm} (\la,\mu)\, = \, \f{ \a(\la) \cdot \a^{-1}(\mu \mp \i c) }{ 2\i \pi (\la-\mu \pm \i c) } \quad \e{with} \quad
\a(\la) \, = \, \exp\bigg\{ \Int{a}{b} \f{ \ln\big[1+F(\mu) \big] }{ \la - \mu } \cdot \f{ \dd \mu }{ 2\i \pi } \bigg\}\;. 
\enq
\end{theorem}
We do remind that the large-$x$ asymptotic behaviour of $\det\big[ \e{id}\, + \, V_0 \big]$ has been obtained in \cite{KozKitMailSlaTerRHPapproachtoSuperSineKernel}.

The paper is organised as follows. In Section \ref{Section RHP initial} we write down the setting of the operator valued Riemann-Hilbert 
problem associated with a one-parameter $t$ deformation of the kernel $V$ given in \eqref{definition noyau integrable V} and prove its unique 
solvability under the assumption of non-vanishing of a Fredholm determinant. 
In Section \ref{Section devpmt elements clef pour NL Steepest Descent}, we discuss an auxiliary scalar operator valued Riemann--Hilbert problem
and implement the first step of the non-linear steepest descent method. Then, in Section \ref{Section parametrices}
we construct the parametrices adapted to out problem what allows us to put the original Riemann--Hilbert problem in correspondence with
 one whose jump matrices are close, in appropriate operator norms, to the identity. 
We then establish the invertibility, in an appropriate functional space, of the singular integral operator 
associated with the last operator valued Riemann--Hilbert problem. 
Finally, in Section \ref{Section calcul asymptotique log det} we build on the Riemann--Hilbert analysis 
so as to prove Theorem \ref{Theorem principal papier}. For the reader's convenience, we gather in the Appendix
certain of the properties of confuent hypergeometric functions that are of interest to our study.

\section{The initial Riemann--Hilbert problem}
\label{Section RHP initial}

\subsection{A few definitions}

We first discuss several notations and conventions that will become handy in the following. 

\vspace{2mm}
 $\bullet $  Throughout the paper, given some oriented curve $\Sg$ in $\Cx$, we agree to denote by $\Ga\big(\Sg \big)$ a small counterclockwise loop around 
$\Sg$.

\vspace{2mm}
 $\bullet $ The superscript $^{\bs{T}}$ will denote the transposition of vectors, \textit{viz}.
\beq
\e{if} \quad \vec{v} \, = \,  \left( \ba{c} v_1 \\ \vdots \\ v_N \ea \right) \qquad \e{then} \quad 
\vec{v}^{\bs{T}} \, = \,  \left(  v_1 \, \dots \,  v_N  \right) \;. 
\enq

\vspace{2mm} $\bullet$ The space $\mc{M}_p(\Cx)$ of $p\times p$ matrices over $\Cx$ is endowed with the norm 
$\big| \big| M \big| \big|=\max_{a,b}|M_{a,b}|$.

\vspace{2mm} $\bullet$ The space $\mc{M}_p\Big(L^2\big( X , \dd \nu \big)\Big)$ denotes the space of $p\times p$ matrix valued
functions on $X$ whose matrix entries belong to $L^2\big( X , \dd \nu \big)$. This space is endowed with the norm 
\beq
\big| \big| M \big| \big|^2_{\mc{M}_p\Big(L^2\big( X , \dd \nu \big)\Big)} \; = \; \Int{X}{} \e{tr}\big[ M^{\dagger}(x)\cdot M(x) \big] \cdot \dd \nu(x) \quad \e{with} \quad 
\Big(M^{\dagger}\Big)_{ab} \, = \, \ov{M_{ba}}
\enq
with $\ov{*}$ being the complex conjugation of $*$.

\vspace{2mm} $\bullet$ $\e{id}$ refers to the identity operator on $L^2(\R^+,\dd s)$, $I_p\otimes \e{id}$ refers to the matrix integral operator on 
$\oplus_{a=1}^{p}L^2\big(\R^+,\dd s \big)$ which has the identity operator on its diagonal.

\vspace{2mm} $\bullet$ Given a vector $\vec{\bs{E}}$ of functions $\bs{E}_a \in L^2(\R^+,\dd s)$
\beq
\vec{\bs{E}}\; = \; \left( \ba{c} \bs{E}_1\\ \vdots\\ \bs{E}_p \ea \right) \quad \e{and} \; \e{a}\; \e{vector} \; \e{of} \; \; 1-\e{forms} \quad 
\vec{\bs{\kappa}} \; = \; \left( \ba{c} \bs{\kappa}_1\\ \vdots\\ \bs{\kappa}_p \ea \right)
\enq
on $L^2(\R^+,\dd s)$, their scalar product refers to the below sum
\beq
\big( \vec{\bs{\kappa}} , \vec{\bs{E}}  \big) \; = \; \sul{a=1}{p} \bs{\kappa}_a[\bs{E}_a]
\enq
in which one evaluates the one-form -appearing to the left- on the function -appearing to the right-. 
Furthermore, the notation $\vec{\bs{E}}\otimes \big(\vec{\bs{\kappa}}\big)^{\bs{T}}$ refers to the matrix operator on 
$\oplus_{a=1}^{p}L^2\big(\R^+,\dd s \big)$ given as 
\beq
\vec{\bs{E}}\otimes \big(\vec{\bs{\kappa}}\big)^{\bs{T}} \; = \; \Big( \bs{E}_{q} \otimes \bs{\kappa}_r \Big)_{q,r=1,\dots,p}
\enq
where $\bs{E}_q\otimes \bs{\kappa}_r$ is the operator on $L^2(\R^+,\dd s)$ acting as
\beq
\big(\bs{E}_q\otimes \bs{\kappa}_r \big)[g] \; = \; \bs{E}_q \cdot \bs{\kappa}_r[g]     \qquad \e{for} \; \e{any} \; g \in L^2(\R^+,\dd s) \;. 
\enq

\begin{defin}

Let $\wh{\Phi}(\la)$ be an integral operator on $\oplus_{a=1}^{p}L^2\big(\R^+,\dd s \big)$ parameterised
 by an auxiliary variable $\la$. Let $\wh{\Phi}(\la\mid s,s^{\prime} )$ denote its $p\times p$ matrix integral kernel.
Given  $\mc{D}$ an open subset of $\Cx$,  we say that  $\wh{\Phi}(\la)$ is a holomorphic in $\la \in \mc{D}$ integral
operator on $\oplus_{a=1}^{p}L^2\big(\R^+,\dd s \big)$ if 
\begin{itemize}
\item point-wise in 
$\big(  s,s^{\prime} \big) \in  \big( \R^+ \big)^2$, the $p\times p$ matrix-valued function 
$\la \mapsto \wh{\Phi}(\la\mid s,s^{\prime} )$ is holomorphic in $\mc{D}$ ; 
\item pointwise in $\la \in \mc{D}$, $\big(  s,s^{\prime} \big) \mapsto \wh{\Phi}(\la\mid s,s^{\prime} ) \in 
\mc{M}_p\Big(L^2\big( \R^+\times \R^+ , \dd s\otimes \dd s^{\prime} \big) \Big)$. 
\end{itemize}

\end{defin}

We also need to define what we mean by $\pm$ boundary values of a holomorphic integral operator. 
There are two kinds of notions that will be of interest for our analysis. On the one hand $L^2$
and on the other hand continuous boundary values. 

\begin{defin}

 Let $\mc{D}$ be an open subset of $\Cx$ and $\Sg_{\Phi}$ an oriented smooth curve in $\Cx$. Let $n(\la)$ be the orthogonal to 
 $\Sg_{\Phi}$ at the point $\la \in \Sg_{\Phi}$. 
 
\noindent We say that a holomorphic in $\la \in \mc{D}\setminus \Sg_{\Phi}$ integral operator $\wh{\Phi}(\la)$ on $\oplus_{a=1}^{p}L^2\big(\R^+,\dd s \big)$ 
admits $L^2$ $\pm$-boundary values $\wh{\Phi}_{\pm}(\la)$ on $ \Sg_{\Phi}$ if
\begin{itemize} 
\item  there exists a matrix valued function $(\la, s, s^{\prime}) \mapsto  \wh{\Phi}_{\pm}(\la\mid s,s^{\prime} )$ belonging to 
$L^2\big( \Sg_{\Phi}\times \R^+\times \R^+ \big)$ and such that 
\beq
\lim_{\eps \tend 0^+}
\big| \big| \wh{\Phi}^{(\pm \eps)} -\wh{\Phi}_{ \pm} \big| \big|_{ \mc{M}_p\Big(L^2\big(  \Sg_{\Phi}\times \R^+\times \R^+ \big)\Big)}  \; = \; 0
\qquad \e{where} \qquad \wh{\Phi}^{( \eps)}(\la\mid s, s^{\prime} ) \; = \; \wh{\Phi}(\la+\eps n(\la) \mid s, s^{\prime} ) \;. 
\nonumber
\enq
the operators $\wh{\Phi}_{\pm}(\la)$ are then defined as the integral operators on  $\oplus_{a=1}^{p}L^2\big(\R^+,\dd s \big)$ 
characterised by the matrix integral kernel $\wh{\Phi}_{\pm}(\la\mid s,s^{\prime} )$. 
\end{itemize}

\noindent We say that a holomorphic in $\la \in \mc{D}\setminus \Sg_{\Phi}$ integral operator $\wh{\Phi}(\la)$ on $\oplus_{a=1}^{p}L^2\big(\R^+,\dd s \big)$ 
admits continuous boundary values $\wh{\Phi}_{\pm}(\la)$ on $\Sg_{\Phi}^{\prime} \subset \Sg_{\Phi}$ if
\begin{itemize}
 \item pointwise in  $\big(  s,s^{\prime} \big) \in  \big( \R^+ \big)^2$ the non-tangential limit
 $\wh{\Phi}(\la\mid s, s^{\prime}) \limit{\la  }{t} \wh{\Phi}_{\pm}(t\mid s,s^{\prime} )$ when $\la$
 approaches $t \in \Sg_{\Phi}^{\prime}$ from the $\pm$ side exists and that 
 the map $t \mapsto \wh{\Phi}_{\pm}(t\mid s,s^{\prime} )$ is continuous on $\Sg_{\Phi}^{\prime}$. 
The operators $\wh{\Phi}_{\pm}(\la)$ are then defined as the integral operators on  $\oplus_{a=1}^{p}L^2\big(\R^+,\dd s \big)$ 
characterised by the matrix integral kernel $\wh{\Phi}_{\pm}(\la\mid s,s^{\prime} )$. 

 \end{itemize}

\end{defin}

 \subsection{The operator-valued Riemann--Hilbert problem}

Let $\la \mapsto \bs{m}_{k}(\la)$ be the below one parameter $t$ family of functions taking values in the space of functions on $\R^+$:
\beq
\bs{m}_{1}(\la)(s) \; \equiv  \, \bs{m}_1(\la;s)\; = \; \sqrt{c} \,\ex{-\f{c s}{2}} \ex{i s t \la}   \qquad \e{and} \qquad
\bs{m}_{2}(\la)(s) \; \equiv  \, \bs{m}_2(\la;s) \; = \; \sqrt{c} \, \ex{-\f{c s}{2}} \ex{-i s t \la}  \;.
\enq
Let $\la \mapsto \bs{\kappa}_{k}(\la)$ be the below one-parameter $t$ family of functions taking values in the space of one-forms on functions on $\R^+$:
\beq
\bs{\kappa}_1(\la)[f]\; = \; \sqrt{c} \Int{0}{+\infty} \ex{-\f{c s}{2}} \ex{ -\i s t \la} f(s) \cdot \dd s   \qquad \e{and} \qquad
\bs{\kappa}_2(\la)[f]\; = \; \sqrt{c} \Int{0}{+\infty} \ex{-\f{c s}{2}} \ex{ \i s t \la } f(s) \cdot \dd s    \; . 
\enq
Note that, uniformly in $\la\in\intff{a}{b}$, $s \mapsto \bs{m}_{k}(\la;s)$ belong to $\big(L^1\cap L^{\infty}\big)(\R^+,\dd s)$ whereas
$\bs{\kappa}_k(\la)$ are one-forms on $L^2(\R^+,\dd s)$. The one-forms and functions introduced above satisfy to 
\beq
\bs{\kappa}_{k}(\la)[\bs{m}_k(\mu)]  \; = \; \f{ \i c \eps_k }{ t(\la-\mu) + \i \eps_k c } \qquad \e{where} \quad  k=1,2 \; \quad 
\e{and} \quad \left\{ \ba{c}   \eps_1 = -1 \\
				    \eps_2 = 1 \ea \right. .
\enq
We are now in position to introduce the vector-valued function $\vec{\bs{E}}_R(\mu)$
and the vector valued one-forms $\vec{\bs{E}}_L(\mu)$:
\beq
\vec{\bs{E}}_L(\mu) \; = \;  F(\mu) \left( \ba{c}   \ex{ - \f{\i x}{2} p(\mu)  } \cdot \bs{\kappa}_{1}(\mu)  \\  
			   -  \ex{  \f{\i x}{2} p(\mu)  }  \cdot \bs{\kappa}_{2}(\mu)  \ea \right)
\qquad \e{and} \qquad 
\vec{\bs{E}}_R(\mu) \; = \; \f{ -1 }{ 2 \i \pi } \left( \ba{c}   \ex{ \f{\i x}{2} p(\mu)  } \cdot \bs{m}_{1}(\mu)  \\ 
								  \ex{ - \f{\i x}{2} p(\mu)  } \cdot \bs{m}_{2}(\mu)   \ea \right) \;. 
\enq
These allow one to construct the integrable integral kernel  $V_t(\la,\mu)$ of the integral operator $V_t$ on $L^2\big( \intff{a}{b} \big)$  as 
\beq
V_t(\la,\mu) \; = \; \f{\Big( \vec{\bs{E}}_L(\la), \vec{\bs{E}}_R(\mu) \Big)    }{\la-\mu}  \; = \; 
\f{ \i c F(\la)   }{ 2\i \pi (\la-\mu) } \cdot \bigg\{ \f{ \ex{ \frac{\i x}{2} [p(\la)-p(\mu)]  } }{ t(\la-\mu) + \i c }
\, + \, \f{  \ex{ \frac{\i x}{2} [p(\mu)-p(\la)]  }  }{ t(\la-\mu) - \i c } \bigg\} \; . 
\enq
Note that the one-parameter $t$ family of integral kernels $V_t(\la, \mu)$ contains the kernel $V(\la,\mu)$
introduced in \eqref{definition noyau integrable V} as a special case; indeed one has $V(\la,\mu)=V_{1}(\la,\mu)$.

\noindent The kernel $V_t(\la,\mu)$ gives rise to the Riemann--Hilbert problem for a $2\times 2$  operator-valued 
matrix $\chi(\la) = I_2\otimes \e{id} + \wh{\chi}(\la) $
\begin{itemize}
 \item $\wh{\chi}(\la)$ is a holomorphic in  $\la \in \Cx \setminus \intff{a}{b}$ integral operator on 
 $L^{2}\big( \R^+, \dd s \big) \oplus L^{2}\big( \R^+, \dd s \big) $;
 \item $\wh{\chi}(\la)$ admits continuous $\pm$-boundary values $\wh{\chi}_{\pm}(\la)$ on $\intoo{a}{b}$; 
\item uniformly in $(s,s^{\prime}) \in \R^+\times \R^+$ and for any compact $K$ such that $\overset{\circ}{K} \supset \{a,b\}$, there exist a constant $C>0$
such that 
\beq
\big| \big| \wh{\chi}\big( \la \mid s,s^{\prime}\big) \big| \big|\; \leq \; \f{ C }{1+|\la|} \cdot \ex{-\f{c}{4} (s+s^{\prime})} \quad \e{on} \quad \Cx \setminus K 
\qquad \e{for} \; \e{some} \; \; C>0\;. 
\label{ecriture bornes chi a l'infini}
\enq

\item there exists $\la$-independent vectors $\vec{\bs{N}}_{\varsigma}$, $\varsigma \in \{a,b\}$ whose entries are functions in 
$\big(L^1\cap L^{\infty}\big)(\R^+,\dd s)$  and an integral operator 
$ \wh{\chi}_{\e{reg}}^{(\vsg)}( \la )$ on $L^{2}\big( \R^+, \dd s \big) \oplus L^{2}\big( \R^+, \dd s \big) $ such that 
\beq
\chi( \la)\; = \; I_2\otimes \e{id} \; + \; \ln \big[ w(\la) \big] \cdot \vec{\bs{N}}_{\varsigma}  \otimes \Big(\vec{\bs{E}}_{L}(\varsigma)\Big)^{\bs{T}}
\; + \; \wh{\chi}_{\e{reg}}^{(\vsg)}( \la ) \qquad \e{where} \quad  w(\la) \, = \,  \f{\la-b}{\la-a} \;. 
\label{ecriture coptmt local chi en a ou b} 
\enq
The integral kernel $\wh{\chi}_{\e{reg}}^{(\vsg)}\big( \la \mid s,s^{\prime}\big)$ satisfies to the bound
\beq
\big| \big|  \wh{\chi}_{\e{reg}}^{(\vsg)}\big( \la \mid s,s^{\prime}\big)  \big| \big| 
\; \leq  \; C \ex{-\f{c}{4} (s+s^{\prime})} (s+1)(s^{\prime}+1) 
\qquad \e{uniformly} \; \e{in} \; \; \la \in U_{\vsg} \;\; \e{and} \; \;  (s,s^{\prime}) \in \R^+\times \R^+
\label{ecriture borne sur chi reg}
\enq
for some open neighbourhood $U_{\vsg}$  of $ \vsg \in \{a,b\}$.

\item the $\pm$ boundary values satisfy $\chi_+(\la) \cdot G_{\chi}(\la) \; = \; \chi_-(\la)$ where the jump matrix reads
\beq
G_{\chi}(\la) \; = \; \left( \ba{cc} \e{id} - F(\la) \cdot  \bs{m}_1(\la)\otimes \bs{\kappa}_{1}(\la)  
									& F(\la) \, \ex{\i x p(\la)} \cdot \bs{m}_1(\la)\otimes \bs{\kappa}_{2}(\la) \\
 -F(\la) \, \ex{- \i x p(\la)} \cdot \bs{m}_2(\la)\otimes \bs{\kappa}_{1}(\la)  
						  & \e{id} + F(\la) \cdot \bs{m}_2(\la)\otimes \bs{\kappa}_{2}(\la)   \ea \right) \;. 
\enq
\end{itemize}

\begin{prop}

The Riemann--Hilbert problem for $\chi$ admits, at most, a unique solutions. 
Furthermore, there exists $\de>0$ and small enough such that for any $t$ such that $|\Im(t)| < \de$
and $\det\big[I+V_t] \not=0$, this unique solution exists and takes the explicit form
\beq
\chi(\la) \;  =  \; 
I_2\otimes \e{id} - \Int{a}{b}  \f{   \vec{\bs{F}}_{R}(\mu) \otimes \Big( \vec{\bs{E}}_{L}(\mu) \Big)^{\bs{T}} }{ \mu- \la }  \cdot \dd \mu    \qquad and \qquad
\chi^{-1}(\la) \; =  \; I_2\otimes \e{id} + \Int{a}{b}   \f{  \vec{\bs{E}}_{R} (\mu)\otimes \Big( \vec{\bs{F}}_{L}(\mu) \Big)^{\bs{T}} }{ \mu - \la }  \cdot \dd \mu 
\label{formules reconstruction chi chi-1 en terms F R et FL}
\enq
where $\vec{\bs{F}}_{R}(\la) $ and $\vec{\bs{F}}_{L}(\la) $ correspond to the solutions to the below linear integral equations
\beq
\vec{\bs{F}}_{R}(\la) \; + \; \Int{a}{b} V_t(\mu,\la) \cdot \vec{\bs{F}}_{R}(\mu)  \cdot \dd \mu \; =\;   \vec{\bs{E}}_{R}(\la)
\qquad and \qquad 
\vec{\bs{F}}_{L}(\la)  \; + \; \Int{a}{b} V_t(\la,\mu) \cdot \vec{\bs{F}}_{L}(\mu)  \cdot \dd \mu  \; =  \;  \vec{\bs{E}}_{L}(\la) \;. 
\enq
The solutions $\vec{\bs{F}}_{R/L}(\la)$ can be constructed in terms of $\chi$ as
\beq
\vec{\bs{F}}_{R}(\mu) \; = \;  \chi(\mu) \cdot \vec{\bs{E}}_{R}(\mu) \qquad and \qquad 
\Big( \vec{\bs{F}}_{L}(\mu) \Big)^{\bs{T}} \; = \;  \Big( \vec{\bs{E}}_{L}(\mu) \Big)^{\bs{T}} \cdot \chi^{-1}(\mu)
\qquad with \quad \la \in \intoo{a}{b} \;. 
\label{reconstruction vecto ops FR et FL}
\enq

\end{prop}

Note that the reconstruction formulae \eqref{reconstruction vecto ops FR et FL} are independent of the $+$ or $-$ boundary values 
of $\chi$ as a consequence of the specific form taken by the jump matrix for $\chi$. 

Furthermore, we do insist that solutions to the Riemann--Hilbert problem for $\chi$ do exist for larger values of $|\Im(t)|$ then those stated in the proposition above.
However, for larger values of $|\Im(t)|$, they define integral operators on weighted $L^2$ spaces $\oplus L^2\big( \R^+, \ex{\a s}\dd s \big)$ 
for some $\a>0$ whose magnitude depends on $|\Im(t)|$. 
Since the conclusions of the above proposition are already enough for the purpose developed in the present paper, we chose not to venture deeper in such 
technicalities.

\Proof

\subsubsection*{$\bullet$ Uniqueness}

For any $\la \in \Cx \setminus \intff{a}{b}$, the matrix-valued operator $\chi(\la)$ decomposes as 
$\chi(\la) \, =\,  I_2 \otimes \e{id} \, + \, \wh{\chi}(\la)$, with an integral kernel 
$\wh{\chi}\big( \la\mid s, s^{\prime} \big)$ that satisfies to \eqref{ecriture bornes chi a l'infini}. This guarantees that its Fredholm determinant 
$\ga(\la) = \det\big[ I_2 \otimes \e{id} \, + \, \wh{\chi}(\la) \big]$ is well defined, \textit{cf}. \cite{GohbergGoldbargKrupnikTracesAndDeterminants}. 
Likewise, it is readily seen by applying Fubbini's and Morera's theorems that
$\ga$ is holomorphic on $\Cx\setminus \intff{a}{b}$. By applying the dominated convergence theorem and the estimates 
\eqref{ecriture bornes chi a l'infini} it is readily seen that $\ga$ admits continuous-boundary values on $\intoo{a}{b}$, which furthermore satisfy
\beq
\ga_{\pm}( \la ) \; =  \;  \det\big[ \chi_{\pm}(\la) \big] \;,
\enq
\textit{ie}. one can exchange the $\pm$ boundary values with the operation of computing the determinant.

We now focus on the behaviour of $\ga$  near the endpoints $a,b$. Starting from \eqref{ecriture coptmt local chi en a ou b} 
one obtains 
\beq
 \det\big[ \chi(\la) \big] \, = \, \det\Big[ I_2\otimes \e{id} \, + \, \wh{\chi}_{\e{reg}}^{(\vsg)}(\la) \Big] 
   \; + \;  \ln \big[ w(\la) \big] \cdot \Big( \vec{\bs{E}}_L(\varsigma), M (\la) \cdot \vec{\bs{N}}_{\varsigma} \Big) 
\quad \vsg \in \{a,b\}
\enq
where $w(\la)$ is as in \eqref{ecriture coptmt local chi en a ou b}. The operator
\beq
M(\la) \; = \;  \lim_{\eta \tend 1} \Bigg\{ \det\Big[ I_2\otimes \e{id} \, + \, \eta \wh{\chi}_{\e{reg}}^{(\vsg)}(\la) \Big]  \cdot  
 \Big( I_2\otimes \e{id} \, + \, \eta \wh{\chi}_{\e{reg}}^{(\vsg)}(\la) \Big)^{-1} \Bigg\} \; = \; I_2\otimes \e{id} \, + \, \wh{M}(\la)
\enq
is well defined even if  $\det\Big[ I_2\otimes \e{id} \, + \,  \wh{\chi}_{\e{reg}}^{(\vsg)}(\la) \Big] = 0$. This can be readily seen from its series of multiple
integral representation, see \textit{eg}. \cite{GohbergGoldbargKrupnikTracesAndDeterminants} and the use of the bounds  \eqref{ecriture borne sur chi reg}. 
The latter ensures that the function 
\beq
 \la \, \mapsto \,  \Big( \vec{\bs{E}}_L(\varsigma), M (\la) \cdot \vec{\bs{N}}_{\varsigma} \Big) 
\enq
is bounded in some open neighbourhood of $\la=\vsg$, hence leading to 
\beq
\big|  \ga(\la) \big| \; \leq \; C \cdot \Big| \ln|\la-a| \cdot \ln|\la-b| \Big| \;  \quad \e{for} \; \e{some} \; \; C>0  
\; \e{and} \; \; \e{when} \; \; \la \tend \varsigma \in \{a,b\}.  
\enq
Finally, independently of $\la \in  \intff{a}{b}$, the integral operator 
$\wh{G}_{\chi}(\la) \, = \, G_{\chi}(\la)- I_2 \otimes \e{id} $ has a $2\times 2$ matrix integral kernels that is smooth 
and such that 
\beq
\big| \big| \wh{G}_{\chi}\big( \la\mid s, s^{\prime} \big) \big| \big| \; \leq \; C \ex{-\f{c (s+s^{\prime}) }{2} } \;. 
\enq
This ensures that the Fredholm determinant of  $G_{\chi}(\la)$ is well defined. 
Then, the multiplicative property of Fredholm determinants along with 
\beq
  \det\Big[ G_{\chi}(\la) \Big]  \, = \, 1  \qquad \e{for} \; \e{any} \quad \la \in \intff{a}{b}
\label{ecriture determinant G chi}
\enq
ensure that $\ga$ solves the scalar Riemann--Hilbert problem
\begin{itemize}
\item $\ga$ is holomorphic on $ \Cx  \setminus \intff{a}{b} $;
\item $\ga$ admits continuous $\pm$-boundary values on $\intoo{a}{b}$ which satisfy $\ga_{+}( \la ) \; = \; \ga_{-}(\la)$;
\item there exists a constant $C>0$ such that when $\la \tend \varsigma \in \{a,b\}$,   $\ga$ satisfies to the bound
\beq
\big|  \ga(\la) \big| \; \leq \; C \cdot \Big| \ln|\la-a| \cdot \ln|\la-b| \Big| \; ; 
\label{ecriture comportement local gamma}
\enq
\item  $\ga(\la) \, = \, 1 \, + \, \e{O}\big(\la^{-1} \big)$ when $\la \tend \infty$ .        
\end{itemize}

The Riemann--Hilbert problem for $\ga$ is uniquely solvable, its solution being  $\ga=1$. 
As a consequence, the matrix-valued operator $\chi(\la)$ is invertible for any $\la \in \Cx \setminus \intff{a}{b}$. Its $\pm$-boundary values 
$\chi_{\pm}(\la)$ are likewise invertible for any $\la \in \intoo{a}{b}$. Assume that $\chi^{(1)}$ and $\chi^{(2)}$ are two solutions to the 
Riemann--Hilbert problem in question.  The operator $G_{\chi}(\la)$ is invertible due to \eqref{ecriture determinant G chi}. 
Therefore, $\Phi = \chi^{(1)} \cdot \big( \chi^{(2)} \big)^{-1} = I_2\otimes \e{id} + \wh{\Phi}$ 
solves a Riemann--Hilbert problem analogous to the one for $\chi$ with the sole exception that 
\begin{itemize}
\item $\Phi_{+}(\la) = \Phi_-(\la)$ on $\intoo{a}{b}$ ;
\item $\Phi(\la)$ admits continuous boundary values on $\intff{a}{b}$; 
\item  $\wh{\Phi}\big(\la\mid s,s^{\prime}\big)$ has, at most, $\e{O}\big( \ln^2|\la-\varsigma| \big)$ singularities at the endpoints 
$\varsigma \in \{a,b\}$ in the sense of \eqref{ecriture coptmt local chi en a ou b} . 
\end{itemize}
 This means that, for any $(s,s^{\prime})\in \R^+\times \R^+$ and lying outside of a set of measure zero, 
the holomorphic matrix-valued functions $\la \mapsto \wh{\Phi}\big( \la \mid s, s^{\prime} \big)$ are 
continous across $\intff{a}{b}$. Being bounded by $0$ at infinity, they are identically zero by Liouville's theorem, 
\textit{viz}.  $\Phi(\la)=I_2\otimes \e{id}$ implying uniqueness.

\subsubsection*{$\bullet$ Existence} 

We chose $\de>0$ and assume the open neighborhood $U$ on which $F$ and $p$ are analytic 
to be relatively compact and small enough so that 
\beq
\big| \ex{\pm \i  s t \la } \big| \; \leq  \; \ex{ \f{c}{4} s } \quad \e{for} \; \e{any} \quad \la \in U\; \; , \;  \de \quad |\Im(t)| \, \leq \, \de 
\quad \e{and} \quad s \in \R^+. 
\label{ecriture bornage des fonction exponentielles}
\enq

We first show that the integral operator defined by \eqref{formules reconstruction chi chi-1 en terms F R et FL} is
indeed a holomorphic in $\la \in \Cx\setminus \intff{a}{b}$ integral operator on $L^2\big(\R^+,\dd s\big) \oplus L^2\big(\R^+,\dd s\big)$. 
Let $R_{t}(\la,\mu)$ be the resolvent  kernel of the inverse operator $\e{id}-\mc{R}_t$ to $\e{id}+\mc{V}_t$. 
This operator exists since $\det\big[ \e{id}+\mc{V}_t\big] \not=0$. Then, one has the representation:
\beq
\vec{\bs{F}}_R(\la;s)\, = \, \vec{\bs{E}}_R(\la;s) \, - \, \Int{a}{b}R_t(\la,\mu) \, \vec{\bs{E}}_R(\mu;s)  \cdot \dd \mu \;.  
\label{ecriture expression explicite pour F right}
\enq
It further follows from \eqref{ecriture bornage des fonction exponentielles} that, 
\beq
\max_{a}\big| \big[ \vec{\bs{E}}_R(\la;s) \big]_a \big| \, \leq \, C \ex{-\f{s c}{4} } \;. 
\enq
 The bounds on $\vec{\bs{E}}_R(\la;s)$ and the regularity of the resolvent kernel $R_t(\la,\mu)$
ensure that 
\beq
\big| \vec{\bs{F}}_R(\la;s) \big| \; \leq \; \ex{- \f{c s }{4 }} \cdot  C \qquad  \e{uniformly} \; \e{in} \; \la \in U \; \; \e{and} \; |\Im(t) | \, \leq \,  \de \;. 
\label{ecriture bornes sur FR}
\enq
Therefore, $\wh{\chi}(\la)$ as defined through \eqref{formules reconstruction chi chi-1 en terms F R et FL} 
does indeed correspond to a holomorphic in $\la \not \in \intff{a}{b}$ integral
operator on  $L^2(\R^+,\dd s) \oplus L^2(\R^+,\dd s)$. 

We now establish the overall bounds \eqref{ecriture bornes chi a l'infini} uniformly away from the endpoints $a$ and  $b$ as well as the local ones 
\eqref{ecriture coptmt local chi en a ou b}-\eqref{ecriture borne sur chi reg} in some
neighbourhood thereof. Since $R_t$ is holomorphic on $U \times U$, one obtains from \eqref{ecriture expression explicite pour F right} that 
\beq
\bigg| \f{ \vec{\bs{F}}_R(\la;s) \cdot \Big(\vec{\bs{E}}_L( \la ; s^{\prime} ) \Big)^{\bs{T}} 
\, - \,  \vec{\bs{F}}_R(\mu;s) \cdot \Big(\vec{\bs{E}}_L(\mu;s^{\prime}) \Big)^{\bs{T}} }{  \la  - \mu }      \bigg| 
\; \leq \; \ex{- \f{c (s+s^{\prime}) }{4 }} \big( 1 + s \big) \big( 1+s^{\prime} \big)  C 
\label{ecriture borne sur noyau integral chi}
\enq
uniformly in $\la, \mu \in U$ and $ |\Im(t) | \, \leq \,  \de $. The latter informations along with the representation 
\bem
\chi(\la) \; =\; \e{id}\otimes I_2 \; - \; \vec{\bs{F}}_R(\la)\otimes \Big( \vec{\bs{E}}_L(\la) \Big)^{\bs{T}}\cdot \ln \big[ w(\la) \big] \\
\, -\, \Int{a}{b} \f{ \vec{\bs{F}}_R(\la;s) \cdot \Big(\vec{\bs{E}}_L( \la ; s^{\prime} ) \Big)^{\bs{T}} 
\, - \,  \vec{\bs{F}}_R(\mu;s) \cdot \Big(\vec{\bs{E}}_L(\mu;s^{\prime}) \Big)^{\bs{T}} }{  \la  - \mu }   \cdot \dd \mu
\end{multline}
ensure that $\wh{\chi}(\la)$ does indeed admit continuous $\pm$-boundary values on $ \intoo{a}{b}$ 
and that it furthermore satisfies to the local \eqref{ecriture coptmt local chi en a ou b}-\eqref{ecriture borne sur chi reg}
and overall \eqref{ecriture bornes chi a l'infini} bounds. 

It now solely remains to prove that $\chi$, as defined through \eqref{formules reconstruction chi chi-1 en terms F R et FL},
does indeed satisfy to the jump condition. 
In fact, this follows from the manipulations outlined in \cite{KorepinSlavnovRHPForNLSMOpertorValuedSettingInto}, 
where the operator valued Riemann--Hilbert problem description of integrable integral operators of $c$-shifted type has been proposed for the first time. 
For the readers convenience, we recall these arguments below. 

It follows directly from the integral representation  \eqref{formules reconstruction chi chi-1 en terms F R et FL} that 
\beq
\chi_+(\la) \, - \, \chi_-(\la) \; = \; -2\i \pi \cdot \vec{\bs{F}}_R(\la) \otimes \Big( \vec{\bs{E}}_L(\la)\Big)^{\bs{T}} \;. 
\enq
Furthermore, by using the explicit expression for $G_{\chi}$, one has that 
\bem
\chi_+(\la)\cdot G_{\chi}(\la) \; = \; \chi_+(\la) \, + \, 
2\i \pi \cdot \vec{\bs{E}}_R(\la) \otimes \Big( \vec{\bs{E}}_L(\la)\Big)^{\bs{T}}
\, - \, 2\i \pi \Int{a}{b} \vec{\bs{F}}_R(\mu) \cdot V_t(\mu,\la) \otimes \Big( \vec{\bs{E}}_L(\la)\Big)^{\bs{T}} \cdot \dd \mu \\
\; = \; \chi_+(\la) \, + \, 2\i \pi \cdot \vec{\bs{F}}_R(\la) \otimes \Big( \vec{\bs{E}}_L(\la)\Big)^{\bs{T}} \;. 
\label{ecriture saut chi avec G chi}
\end{multline}
where, in the last equality, we have used the integral equation satisfied by $\vec{\bs{F}}_R(\la)$. By using the above 
two relations, one indeed obtains that $\chi$ satisfies to the jump conditions. 
Finally, it follows from the first equality in \eqref{ecriture saut chi avec G chi} that 
\beq
\chi_+(\la) \cdot \vec{\bs{E}}_R(\la) \otimes \Big( \vec{\bs{E}}_L(\la)\Big)^{\bs{T}}  \; = \; 
\vec{\bs{F}}_R(\la) \otimes  \Big( \vec{\bs{E}}_L(\la)\Big)^{\bs{T}} \;. 
\enq
Acting with both sides of this equality on a vector function $\vec{G}$ such that $\big(  \vec{\bs{E}}_L(\la) , \vec{G}\big)\not=0$
for $\la \in \intff{a}{b}$, we obtain \eqref{reconstruction vecto ops FR et FL}. 
The proofs of similar statements relative to $\chi^{-1}$ are left to the reader. \qed

\vspace{4mm}

We remind that it is a classical fact \cite{ItsIzerginKorepinSlavnovDifferentialeqnsforCorrelationfunctions} that 
the resolvent operator $\mc{R}_t$ to $\mc{V}_{t}$ belongs to the class of integrable
integral operator and that its integral kernel $R_{t}(\la,\mu)$ reads 
\beq
R_t(\la,\mu) \; = \; \f{ \Big(  \vec{\bs{F}}_L(\la) , \vec{\bs{F}}_R(\mu) \Big) }{ \la- \mu } \;. 
\label{ecriture formule pour noyau resolvent}
\enq
%
%
%




\section{Towards the implementation of the non-linear steepest descent method}
\label{Section devpmt elements clef pour NL Steepest Descent}

\subsection{Auxiliary operator-valued  scalar Riemann--Hilbert problems}
\label{SoussectionAuxiliaryScalarRHP}
Let
\beq
\tau_{1}(\la) \; = \;  -\f{ F(\la) }{ 1 + F(\la) }  \qquad \e{and} \qquad 
\tau_{2}(\la) \; = \; F(\la) \;. 
\enq
In the present section, we investigate the solution of two operator-valued scalar Riemann--Hilbert problems that will become useful 
in our future handlings. Before stating the Riemann--Hilbert problems of interst, we however need
to introduce a function that will become handy:
\beq
\nu(\la) \; = \; \f{ -1 }{ 2 \i \pi } \cdot \ln \big[ 1\, + \, F(\la) \big]
\qquad \e{and} \qquad \a(\la) \; = \; \exp\bigg\{ \Int{a}{b}  \f{ \nu(\mu) }{ \mu-\la } \cdot \dd \mu \bigg\}
\enq
The Riemann--Hilbert problem for $\be_{k}=\e{id}+\wh{\be}_{k}$ with $k=1,2$ reads:
\begin{itemize}
 \item $\wh{\be}_k(\la)$ is a holomorphic in $\la \in \Cx \setminus \intff{a}{b}$ integral operator on $L^{2}\big( \R^+, \dd s \big)$;
 \item $\wh{\be}_k(\la)$ admits continuous $\pm$-boundary values $\wh{\be}_{k;\pm}$ on $\intoo{a}{b}$;
\item uniformly in $(s,s^{\prime}) \in \R^+\times \R^+$ and for any compact $K$ such that $\e{Int}(K)\supset \{ a , b \}$, there exist a constant $C>0$
such that 
\beq
\big| \wh{\be}_k\big( \la \mid s,s^{\prime}\big) \big| \; \leq \; \f{ C }{1+|\la|} \cdot \ex{-\f{c}{4} (s+s^{\prime})} \quad \e{for} \quad \Cx \setminus K \;. 
\label{ecriture bornage a infini noyau beta k}
\enq
\item There exists a function $\bs{n}_{k;\varsigma} \in \big(L^1\cap L^{\infty}\big) \big(\R^+,\dd s \big)$ 
and a neighbourhood $U_{\vsg}$ of $\varsigma \in \{a,b\}$ such that for $\la$ in 
\beq
\wh{\be}_k(\la) \; = \; \big[ w(\la) \big]^{ -\eps_k \nu (\vsg)  } \cdot \bs{n}_{k;\varsigma}\otimes \bs{\kappa}_{k}(\varsigma) 
\;+ \; \wh{\be}_{k;\e{reg}}^{(\vsg)}(\la)
\enq
where $w(\la)$ is as given in \eqref{ecriture coptmt local chi en a ou b} while, for any $\la \in U_{\vsg}$,
\beq
\big| \wh{\be}_{k;\e{reg}}^{(\vsg)}\big( \la \mid s, s^{\prime} \big) \big| \; \leq \; C \ex{-\f{c}{4}(s+s^{\prime})}(s+1)(s^{\prime}+1) \quad \e{for} \; \e{some} \quad C>0 \;. 
\enq
\item the boundary values satisfy $\be_{k;+}(\la) \cdot \Big( \e{id} \, + \, \tau_{k}(\la) \cdot \bs{m}_{k}(\la)\otimes \bs{\kappa}_{k}(\la) \Big) \; = \; \be_{k;-}(\la)$. 
\end{itemize}

\begin{prop}
 
There exists $\de>0$ small enough such that the Riemann--Hilbert problem for $\be_{k}$ admits a unique solution provided that $1+\tau_{k}(\la) \not=0$ on $\intff{a}{b}$
and $|\Im(t)|<\de$. Furthermore, the solution exists as soon as 
\beq
|\Im(t)|<\de \quad  \e{and} \quad   \det_{\Ga(\intff{a}{b}) } \big[ \e{id} \, + \, \mc{U}_{k;t} \big] \; \not= \; 0
\enq
where the integral kernel $U_{k;t}(\la,\mu)$ of the integral operator $\mc{U}_{k;t}$ acting on $L^2\big( \Ga(\intff{a}{b}) \big)$ reads 
\beq
U_{k;t}(\la,\mu) \; = \;   -  t \f{  \a_{k}(\la) \cdot \a_k^{-1}(\mu + \i \eps_k \tf{c}{t})  }{ 2 \i \pi \cdot \big[ t (\mu-\la) + \i \eps_k c \big]  } \; 
\qquad \e{with} \quad \eps_1=-1 \;\;  \e{and}  \; \; \eps_2=1 \; , 
\label{definition noyau integral U k et t}
\enq
in which 
\beq
\a_{k}(\la)\; = \; \exp\bigg\{  \Int{a}{b} \f{ \nu_k(\mu)  }{ \mu - \la } \cdot \dd \mu  \bigg\}  \qquad \e{with}  \qquad
\nu_{k}(\mu) \; = \; \f{-1}{2\i \pi} \ln \big[ 1+\tau_k(\mu) \big] \;. 
\label{definition solution alpha k}
\enq

\end{prop}

Note that one has 
\beq
\nu_k(\la) \, = \, \eps_k \nu(\la) \qquad \e{and} \qquad \a_k(\la) \, = \, \big[ \a(\la) \big]^{\eps_k} \; .
\enq

\Proof 

\subsubsection*{$\bullet$ Uniqueness}

For any $\la \in \Cx \setminus \intff{a}{b}$, in virtue of \eqref{ecriture bornage a infini noyau beta k} the Fredholm determinant
$\a_k(\la) = \det\big[ \e{id} \, + \, \wh{\be}_k(\la) \big]$ is well defined. 
It follows from the reasoning outlined previously that $\a_k$ is holomorphic on $\Cx\setminus \intff{a}{b}$ and that 
it admits continuous $\pm$-boundary values on $\intoo{a}{b}$, which furthermore satisfy
\beq
\a_{k;\pm}( \la ) \; =  \;  \det\big[ \be_{k;\pm}(\la) \big] \;. 
\enq
Since the integral kernel of $\tau_k(\la)\cdot \bs{m}_k(\la) \otimes \bs{\kappa}_k(\la)$ is bounded by $C\ex{-\f{c}{4}(s+s^{\prime})}$, 
this independently of $\la \in  \intff{a}{b}$, the multiplicative property of Fredholm determinants and 
the local structure of $\wh{\be}_{k}$ in some neighbourhood of the endpoints $a,b$
ensure that $\a_{k}$ solves the scalar Riemann--Hilbert problem
\begin{itemize}
\item $\a_k$ is holomorphic on  $ \Cx  \setminus \intff{a}{b}$;
\item $\a_k$ admits continuous $\pm$ boundary values $\a_{k;\pm}$ on $\intoo{a}{b}$ which satisfy $\a_{k;+}( \la )\cdot \Big( 1+\tau_k(\la) \Big) \; = \; \a_{k;-}(\la)$;
\item $\a_k(\la)  = \e{O} \Big( |w(\la)^{- \eps_k \nu(\varsigma) } |  \Big)$ when $\la \tend \varsigma \in \{a,b\}$;
\item  $\a_k\, = \, 1 \, + \, \e{O}\big(\la^{-1} \big)$ when $\la \tend \infty$ .        
\end{itemize}
The hypothesis of the theorem ensure the unique solvability of this scalar problem, with its solution being given by \eqref{definition solution alpha k}. 
In particular, $\a_k(\la)\not=0$ for $\la \in \Cx \setminus \intff{a}{b}$, just as  $\a_{k;\pm}(\la)\not=0$ for $\la \in \intoo{a}{b}$. As a consequence, 
 the operator $\be_k(\la)$ is invertible for any $\la \in \Cx \setminus \intff{a}{b}$. Furthermore, its $\pm$-boundary values 
$\be_{k;\pm}(\la)$ are also invertible for any $\la \in \intoo{a}{b}$. Assume that $\be_k^{(1)}$ and $\be_k^{(2)}$ are two solutions to the 
Riemann--Hilbert problem in question. Observe that, due to 
\beq
\det\Big[\e{id} \, + \, \tau_k(\la)\cdot \bs{m}_k(\la) \otimes \bs{\kappa}_k(\la) \Big] \, =\,  1+\tau_k(\la)  \not= 0 \; , 
\enq
the operator $\e{id} \, + \, \tau_k(\la)\cdot \bs{m}_k(\la) \otimes \bs{\kappa}_k(\la)$ is invertible. As a consequence, 
$\ga_k= \be_k^{(1)} \cdot \big( \be_k^{(2)} \big)^{-1}$ solves a Riemann--Hilbert problem analogous to the one for $\be_k$ with the sole exception that 
now $\ga_{k;+}(\la) = \ga_{k;-}(\la)$ on $\intoo{a}{b}$ and that $\wh{\ga}_{k}\big(\la,\mid s,s^{\prime} \big) $ exhibits at most 
$\e{O} \Big( |w(\la)^{-2\eps_{k}\nu(\varsigma)}|  \Big)$ singularities in $\la$ when $\la \tend \varsigma \in \{a,b\}$. 
This means that, for any $(s,s^{\prime})\in \R^+\times \R^+$, 
the holomorphic function $\la \mapsto \wh{\ga}\big( \la \mid s, s^{\prime} \big)$ is continuous across $\intoo{a}{b}$ 
and has removable singularities at the endpoints. This function is thus entire and, being bounded by $0$ at infinity, it is identically zero by Liouville's 
theorem, \textit{viz}.  $\ga_k(\la)=\e{id}$.

\subsubsection*{$\bullet$ Existence}

Due to its unique solvability, the solution to the Riemann--Hilbert problem for $\be_k$, if it exists, is in one-to-one correspondence with the solution to 
the singular integral equation \cite{BealsCoifmanScatteringInFirstOrderSystemsEquivalenceRHPSingIntEqnMention}
\beq
\be_{k;+}(\la) \; = \;  \e{id} \; - \; \mc{C}_+\big[ \be_{k;+} \cdot \tau_{k} \cdot \bs{m}_{k}\otimes  \bs{\kappa}_{k} \big](\la) 
\quad \e{where} \quad \mc{C}[f]( \la ) \; = \; \Int{a}{b} \f{ f(\mu) }{ \mu - \la } \cdot \f{ \dd \mu }{2 \i \pi}  
\enq
and $\mc{C}_{+}[f]( \la )$ stands for the $+$ boundary value of $\mc{C}[f]( \la )$ on $\intoo{a}{b}$.
More precisely, the solution $\be_{k}$ can be represented as
\beq
\be_{k}(\la) \; = \;  \e{id} \; - \; \mc{C}\big[ \be_{k;+} \cdot \tau_{k} \cdot \bs{m}_{k}\otimes  \bs{\kappa}_{k} \big](\la) \;. 
\enq
We transform the singular integral equation for $\be_{k;+}$ into one for the function
\beq
\rho_k(\la;s) \; = \; \Big( \be_{k;+}(\la)\cdot \bs{m}_k(\la) \Big)(s) \;. 
\enq
We obtain
\beq
\rho_k(\la;s) \; = \; h_k(\la;s) \, - \, \mc{C}_+\big[ \tau_k(*) \rho_k(*;s) \big](\la)
\quad \e{where} \quad
 h_k(\la;s) \; = \; \sqrt{c} \ex{-\f{c s }{2} - \eps_k \i t s \la}
 \, + \, \Int{a}{b} \f{ t \tau_k(\mu) \rho_k(\mu;s)  }{ t(\mu-\la) + \i  \eps_k  c  } \cdot \f{ \dd \mu }{ 2 \i \pi } \;. 
\label{ecriture eqn pour fct rho}
\enq
Above the $*$ indicates the running variable of the function on which the Cauchy transform acts and we remind that 
$\eps_1=-1$ while $\eps_2=1$.
Equation \eqref{ecriture eqn pour fct rho} can be recast as a non-homogeneous Riemann--Hilbert problem 
for the function
\beq
\aleph(\la;s) \; = \;  \Int{a}{b} \f{ \tau_k(\mu) \rho_k(\mu;s)  }{ \mu-\la } \cdot \f{ \dd \mu }{ 2 \i \pi } \;. 
\enq
Indeed, $\la \mapsto \aleph(\la\mid s) $ is  holomorphic on $\Cx \setminus \intff{a}{b}$, decays as $\e{O}\big(\la^{-1} \big)$ and satisfies to 
the non-homogeneous jump conditions
\beq
\aleph_+(\la;s)\cdot \big( 1+\tau_k(\la) \big) \, - \,  \aleph_-(\la;s) \; = \; 
\tau_k(\la) \cdot h_k(\la;s) \;. 
\enq
This non-homogeneous Riemann-Hilbert problem  is readily solved by standard techniques \cite{MuskhelishviliSingularIntegralEquations} leading to 
\beq
\aleph(\la;s) \; = \; \a_k(\la) \cdot \Int{a}{b}  \f{ \a_{k;-}^{-1}(\mu)  }{ \mu-\la } \cdot \tau_k(\mu) h_k(\mu;s)  \cdot \f{ \dd \mu }{ 2 \i \pi }  \; . 
\enq
We do stress that the functions $\a_k$ are well defined as a consequence of our hypothesis on $F$. 
Making most of the expression for $h_k$, one gets 
\beq
\aleph(\la;s)\; = \; h_k(\la;s) \, - \, \a_k(\la) \hspace{-2mm} \Oint{ \Ga( \intff{a}{b})  }{} \hspace{-3mm} 
 \f{ \sqrt{c} \cdot \ex{-\f{c s }{2} - \i \eps_k t s \mu} }{ \a_k(\mu) \cdot ( \mu - \la )  } \cdot \f{ \dd \mu }{ 2 \i \pi }
\, - \, 
\a_k(\la) \Int{ a }{ b }  \f{ t \cdot \a_k^{-1}(\mu + \tf{ \i \eps_k c}{t} )  }{ t(\mu-\la) + \i \eps_k c  } \cdot \tau_k(\mu) \rho_k(\mu;s)\cdot \f{ \dd \mu }{ 2 \i \pi } 
\label{representation alternative pour aleph}
\enq
for $\la$ belonging to a small vicinity of $\intff{a}{b}$. We remind that, in \eqref{representation alternative pour aleph}, $\Ga\big( \intff{a}{b}\big) $
stands for a small counterclockwise loop around the segment $\intff{a}{b}$ and the point $\la$.  As a consequence, $\rho_k$ solves the linear integral equation 
\beq
\Big( \e{id} \, + \, \mc{K}_{k;t} \Big)[\rho_k(*;s)](\la) \; = \; \a_{k;+}(\la) 
\Oint{ \Ga\big( \intff{a}{b}\big)  }{}  \f{ \sqrt{c} \cdot  \ex{-\f{c s }{2} - \i \eps_k t s \mu} }
{ \a_k(\mu) \cdot ( \mu-\la ) } \cdot \f{ \dd \mu }{ 2 \i \pi }
\label{eqn integrale pour fct rhok} 
\enq
where the integral kernel $K_{k;t}(\la,\mu)$ of the integral operator $\mc{K}_{k;t}$ on $L^2\big( \intff{a}{b} \big)$ reads
\beq
K_{k;t}(\la,\mu) \; = \; -    t \f{ \a_{k;+}(\la) \cdot \a_k^{-1}(\mu + \i \eps_k \tf{c}{t} )  }{ 2 \i \pi \cdot \big( t(\mu-\la) + \i \eps_k c \big)  } \cdot \tau_k(\mu) \;.
\enq
In fact, using the jump condition satisfied by $\a_k$ in the form $\a_{k;-}-\a_{k;+} \, = \, \a_{k;+} \tau_k$, one can recast the kernel as 
\beq
K_{k;t}(\la,\mu) \; = \; -  t \Big( \a_{k;-}(\mu) - \a_{k;+}(\mu) \Big) \cdot \f{ \a_{k;+}(\la) }{  \a_{k;+}(\mu) } \cdot 
						      \f{  \a_k^{-1}(\mu + \i \eps_k \tf{c}{t})  }{ 2 \i \pi \cdot \big( t(\mu-\la) + \i \eps_k c )  } \;. 
\enq
As a consequence, one gets that 
\beq
\det_{\intff{a}{b}}\big[ \e{id} \, + \, \mc{K}_{k;t} \big] \; = \;  \det_{\Ga(\intff{a}{b})}\big[ \e{id} \, + \, \mc{U}_{k;t} \big]
\enq
where the integral kernel $U_{k;t}(\la,\mu)$ of the integral operator $\mc{U}_{k;t}$ acting on $L^2\big( \Ga( \intff{a}{b} ) \big)$ is as defined in 
\eqref{definition noyau integral U k et t}. 
The operator $\e{id}+\mc{K}_{k,t}$ is thus invertible. Let $\bs{\rho}_{k}(\la)$ denote the function $\Big(\bs{\rho}_{k}(\la)\Big)(s) \, = \, \rho_{k}(\la;s)$
where $\rho_{k}(\la;s)$ is as defined by \eqref{eqn integrale pour fct rhok}. 
As a consequence, 
\beq
\be_k(\la) \; = \; \e{id} \, - \, \Int{a}{b} \f{ \tau_k(\mu) \, \bs{\rho}_k(\mu)\otimes \bs{\kappa}_k(\mu) }{ \mu - \la } \cdot \f{ \dd \mu }{ 2\i \pi }
\enq
is the good candidate for the unique solution to the Riemann--Hilbert problem for $\be_k$. 
It is readily checked by repeating the arguments invoked in the proof of the existence of solutions to the Riemann--Hilbert problem for $ \chi$,
that $\be_k$ as defined above does satisfy all the requirements stated in the Riemann--Hilbert problem for $\be_k$. 
\qed

\subsection{A regularity lemma}

In the analysis that will follow, there will arise the one-parameter $\la$ integral operator on $L^2\big(\R^+, \dd s ) \oplus L^2\big(\R^+, \dd s )$ defined as 
\beq
\bs{O}(\la) \; = \; \left( \ba{cc}  \be_{1}(\la) \cdot \bs{m}_1(\la) \otimes   \bs{\kappa}_1(\la) \cdot \be_{1}^{-1}(\la)   & 
					    \a^2(\la)  \be_{1}(\la)\cdot  \bs{m}_1(\la) \otimes   \bs{\kappa}_2(\la) \cdot \be_{2}^{-1}(\la)  \\
		\a^{-2}(\la)  \be_{2}(\la) \cdot  \bs{m}_2(\la) \otimes   \bs{\kappa}_1(\la) \cdot  \be_{1}^{-1}(\la)   &
					\be_{2}(\la) \cdot  \bs{m}_2(\la) \otimes   \bs{\kappa}_2(\la) \cdot  \be_{2}^{-1}(\la)   \ea \right) \;. 
\label{definition operateur regulier O}
\enq
The main point is that even though the individual operators appearing in its matrix elements have cuts, the operator, as a whole, 
is regular. More precisely, one has the

\begin{lemme}
\label{Lemme regularite certains op a trace}
There exists an open neighbourhood $V$ of the segment $\intff{a}{b}$ such that the integral operator $\bs{O}(\la)$ on $L^2\big(\R^+, \dd s ) \oplus L^2\big(\R^+, \dd s )$ 
defined in \eqref{definition operateur regulier O} is holomorphic on $V$. 
\end{lemme}
\Proof

By composition of holomorphic operators, $\bs{O}$ is holomorphic in $V\setminus \intff{a}{b}$, with $V$ a sufficiently small open neighbourhood of $\intff{a}{b}$. 
We thus need to show that it is continuous across $\intff{a}{b}$ and that it has removable singularities at $a,b$. 
For this purpose, observe that 
\beq
\Big( \e{id} \, + \, \tau_k(\la) \cdot \bs{m}_k(\la) \otimes \bs{\kappa}_k(\la) \Big) \cdot 
			\Big( \e{id} \, - \, \f{ \tau_k(\la) }{ 1+ \tau_k(\la) } \cdot \bs{m}_k(\la) \otimes \bs{\kappa}_k(\la) \Big) \; = \; \e{id} \;. 
\enq
Hence, since $\be_{k;\pm}(\la)$ are invertible for all $\la \in \intoo{a}{b}$, one has 
\beq
\be_{k;-}^{-1}(\la) \; = \; \Big( \e{id} \, - \, \f{ \tau_k(\la) }{ 1+ \tau_k(\la) } \bs{m}_k(\la) \otimes \bs{\kappa}_k(\la) \Big) \cdot \be_{k;+}^{-1}(\la) \;. 
\enq
As a consequence, one obtains the jump conditions
\beq
\be_{k;+}(\la) \cdot \bs{m}_k(\la)  \; = \; \be_{k;-}(\la) \cdot \bs{m}_k(\la) \cdot \f{ 1 }{   1+ \tau_k(\la)   } \qquad \e{and} \quad 
\bs{\kappa}_k(\la) \cdot  \be_{k;+}^{-1}(\la) \; = \;  \bs{\kappa}_k(\la) \cdot \be_{k;-}^{-1}(\la) \cdot  \big( 1+ \tau_k(\la)  \big)
\enq

These are enough so as to conclude that $\bs{O}\big(\la\mid s, s^{\prime}\big)$ is continuous across $\intoo{a}{b}$.
It also has removable singularities at the endpoints as readily inferred from the local behaviour of $\a$
and of the operators $\be_{k}$ around $a$ or $b$. It thus extends to a holomorphic function in 
some open neighbourhood of $\intff{a}{b}$. \qed

\subsection{Asymptotic resolution of the Riemann--Hilbert problem for $\chi$}

Observe that one has the factorisation
\bem
G_{\chi}(\la) \; = \;  \left( \ba{cc} \e{id}  &  \f{F(\la) \ex{\i x p(\la)} }{1+F(\la) } \bs{m}_1(\la)\otimes \bs{\kappa}_{2}(\la) \\
 0 & \e{id}    \ea \right)
\cdot \left( \ba{cc} \e{id} -  \f{F(\la) }{1+F(\la) } \cdot \bs{m}_1(\la)\otimes \bs{\kappa}_{1}(\la)  & 0 \\
0 & \e{id} + F(\la) \bs{m}_2(\la)\otimes \bs{\kappa}_{2}(\la)   \ea \right)  \\
\times \left( \ba{cc} \e{id}  & 0 \\
-  \f{F(\la)   \ex{- \i x p(\la)} }{1+F(\la) } \bs{m}_2(\la)\otimes \bs{\kappa}_{1}(\la)  & \e{id}   \ea \right) \;. 
\end{multline}
One can factor the diagonal operator valued matrix appearing in the centre by using the solutions of the operator valued scalar Riemann--Hilbert problems
considered in Section \ref{SoussectionAuxiliaryScalarRHP}.
This allows one to factorise the jump matrix $G_{\chi}$ as 
\beq
G_{\chi}(\la) \; = \;  \left( \ba{cc} \be_{1;+}^{-1}(\la) & 0  \\ 
						0  &  \be_{2;+}^{-1}(\la) 	\ea \right)
\cdot M_{\ua;+}(\la) \cdot M_{\da;-}(\la) \cdot \left( \ba{cc} \be_{1;-}(\la) & 0  \\ 
						0  &  \be_{2;-}(\la) 	\ea \right)
\enq
where the matrices $M_{\ua/\da}$ read
\beq
M_{\ua}(\la) \; = \; \left( \ba{cc} \e{id}  &  \bs{P}(\la)  \ex{\i x p(\la)} \\
		      0  & \e{id}  \ea \right)   \qquad  \e{and} \qquad 
M_{\da}(\la) \; = \; \left( \ba{cc} \e{id}  &  0 \\
		      \bs{Q}(\la)  \ex{- \i x p(\la)}  & \e{id}  \ea \right) 
\enq
in which 
\beq
\bs{P}(\la) \; = \;   \f{ F(\la) }{ 1+F(\la)  }  \be_{1}(\la) \cdot \bs{m}_1(\la)\otimes \bs{\kappa}_{2}(\la) \cdot \be_{2}^{-1}(\la)
 \qquad \e{and} \qquad 
\bs{Q}(\la) \; = \;  - \f{ F(\la) }{ 1+F(\la)  }  \be_{2}(\la) \cdot  \bs{m}_2(\la)\otimes \bs{\kappa}_{1}(\la) \cdot \be_{1}^{-1}(\la) \;. 
\enq
Note that the operators $\bs{P}$ and $\bs{Q}$ can be recast as 
\beq
\bs{P}(\la) \; = \; -2 \i \ex{ \i\pi \nu(\la) } \f{ \sin\big[ \pi \nu(\la) \big] }{ \a^2(\la)  } \cdot \bs{O}_{12}(\la) \qquad \e{and} \qquad
\bs{Q}(\la) \; = \; 2 \i \ex{ \i \pi \nu(\la) } \sin\big[ \pi \nu(\la) \big]  \a^2(\la)  \cdot   \bs{O}_{21}(\la)
\label{ecriture represebtation avec sing expl pour P et Q}
\enq
where $ \bs{O}(\la)$ is as defined by \eqref{definition operateur regulier O}. 

Thus, agreeing to denote 
\beq
\Xi(\la) \; = \; \chi(\la) \cdot \left( \ba{cc}   \be_1^{-1}(\la) &  0  \\ 
						      0   & \be_2^{-1}(\la) \ea \right)
\enq
and then defining the matrix $\Ups$ and the contour $\Sg_{\Ups}$ according to Fig.~\ref{contour pour le RHP de Y} one gets, 
upon repeating the steps already explained previously, that $\Ups(\la) = I_2\otimes \e{id} + \wh{\Ups}(\la) $ solves the Riemann--Hilbert problem
\begin{itemize}
 \item $\wh{\Ups}(\la)$ is a holomorphic in  $\la \in \Cx \setminus \Sg_{\Ups}$ integral  operator on $L^{2}\big( \R^+, \dd s \big) \oplus L^{2}\big( \R^+, \dd s \big) $;
 \item $\wh{\Ups}(\la)$ admits continuous $\pm$-boundary values $\wh{\Ups}_{\pm}(\la)$  on $\Sg_{\Ups}\setminus \{a,b\}$;
 \item uniformly in $(s,s^{\prime}) \in \R^+\times \R^+$ and for any compact $K$ such that $\e{Int}(K) \supset \{a,b\}$, 
there exist a constant $C>0$ such that 
\beq
\big| \wh{\Ups}\big( \la \mid s,s^{\prime}\big) \big| \; \leq \; \f{ C }{1+|\la|} \cdot \ex{-\f{c}{4} (s+s^{\prime})} \quad \e{for} \quad \Cx \setminus K \;. 
\enq
\item there exists an open neighbourhood $U_{\vsg}$ of $\vsg \in \{a,b\}$,   
vector valued functions $\vec{\bs{N}}_{\varsigma}$ as well as functions $\wt{\bs{n}}_{k;\varsigma}$, $k=1,2$,  
all belonging to $\big(L^{1}\cap L^{\infty}\big)\big( \R^+, \dd s\big)$ such that, for $\la \in U_{\varsigma} \cap H_{III}$ 
one has $\Ups(\la) \; = \; \Ups_{H_{III}}(\la)$ where 
\bem
\Ups_{H_{III}}(\la) \; = \; \Big( I_2 \otimes \e{id} \; + \; \ln\big[ w(\la) \big] \cdot \vec{\bs{N}}_{\varsigma}\otimes \big( \vec{\bs{E}}_{L}(\varsigma) \big)^{\bs{T}} \; + \; 
\wh{R}_{ \Ups}^{(\varsigma)}(\la) \Big)    \\ 
\times \; \left(\ba{cc} 
\e{id} + \big[ w(\la) \big]^{\nu_{1}(\la)}  \wt{\bs{n}}_{1;\varsigma}\otimes \bs{\kappa}_1(\varsigma)  \; + \; r_{1;\Ups}^{(\varsigma)}(\la) & 0  \\
0 & \e{id} + \big[ w(\la) \big]^{\nu_{2}(\la)}  \wt{\bs{n}}_{2;\varsigma}\otimes \bs{\kappa}_1(\varsigma)  \; + \; r_{2;\Ups}^{(\varsigma)}(\la) \ea \right) \;,
\nonumber
\end{multline}
$w(\la)$ is as defined in \eqref{ecriture coptmt local chi en a ou b}, and $\wh{R}_{ \Ups}^{(\varsigma)}(\la)$, resp. $r_{k;\Ups}^{(\varsigma)} $, 
is an integral operator on $L^2(\R^+,\dd s)\oplus L^2(\R^+,\dd s) $, 
resp. $L^2(\R^+,\dd s)$, such that for any $\la \in U_{\vsg}$ 
\beq
\big| \big| \wh{R}_{ \Ups}^{(\varsigma)}(\la\mid s, s^{\prime})  \big| \big| \; \leq  \;  C
\ex{-\f{c}{4}(s+s^{\prime})} (s+1)(s^{\prime}+1)
\qquad \e{resp.}  \qquad 
\big| r_{k;\Ups}^{(\varsigma)}(\la \mid s, s^{\prime}) \big| \; \leq  \; C \ex{-\f{c}{4}(s+s^{\prime})}(s+1)(s^{\prime}+1)  
\enq
for some constant $C>0$. Furthermore, one has that 
\beqa
\Ups(\la) & = &  \Ups_{H_{III}}(\la) \cdot \left( \ba{cc} \e{id} & \big[ w(\la) \big]^{-2\nu(\la) } \bs{P}_{\e{reg}}(\la) \\ 
						  0     & \e{id}  \ea \right)   \quad \e{where} \quad  \la \tend \varsigma \in \{a,b\}
\quad \e{with} \quad \la \in U_{\varsigma} \cap H_{I}  \nonumber\\ 
\Ups(\la) & = &  \Ups_{H_{III}}(\la) \cdot \left( \ba{cc} \e{id} & 0 \\ 
						  \big[ w(\la) \big]^{2\nu(\la) } \bs{Q}_{\e{reg}}(\la)     & \e{id}  \ea \right)   \qquad \e{where} \quad  \la \tend \varsigma \in \{a,b\}
\quad \e{with} \quad \la \in U_{\varsigma} \cap \in H_{II} \nonumber 
\eeqa
where $\bs{P}_{\e{reg}}(\la)$ and $\bs{Q}_{\e{reg}}(\la) $ are integral operators on $L^2(\R^+,\dd s)$ such that, 
\beq
\big| \bs{P}_{\e{reg}}(\la\mid s, s^{\prime} ) \big|  \; \leq \;  C \ex{-\f{c}{4}(s+s^{\prime})} (s+1)(s^{\prime}+1) 
\qquad \e{and} \qquad
\big|  \bs{Q}_{\e{reg}}(\la\mid s, s^{\prime} ) \big|  \; \leq  \;  C \ex{-\f{c}{4}(s+s^{\prime})} (s+1)(s^{\prime}+1) 
\label{ecriture proprietes operateurs P et Q reg}
\enq
for some constant $C>0$ and any $\la \in U_{\vsg}$.  

\item the boundary values satisfy $\Ups_+(\la) G_{\Ups}(\la) \; = \; \Ups_-(\la)$ where the jump matrix reads
\beq
G_{\Ups}(\la) \; = \; M_{\ua}(\la) \quad \e{for} \quad \la \in \Ga_{\ua} \qquad \e{and} \qquad 
G_{\Ups}(\la) \; = \; M_{\da}^{-1}(\la) \quad \e{for} \quad \la \in \Ga_{\da} \; . 
\enq
\end{itemize}

Again, this Riemann--Hilbert problem is uniquely solvable and hence, its solution is in one-to-one correspondence with the one 
to the Riemann--Hilbert problem for $\chi$. The fact that the operators $\bs{P}_{\e{reg}}(\la)$ and $\bs{Q}_{\e{reg}}(\la)$
satisfy \eqref{ecriture proprietes operateurs P et Q reg} follows from \eqref{ecriture represebtation avec sing expl pour P et Q}, 
Lemma \ref{Lemme regularite certains op a trace} as well as from the local 
behaviour of $\a$ around $\la=\vsg \in \{a,b\}$. Finally, the local behaviour 
of $\Ups$ around $\vsg\in \{a,b\}$ is inferred from the one of $\chi$, \textit{cf}.
Fig. \ref{contour pour le RHP de Y}.

\begin{figure}[h]
\begin{center}

\begin{pspicture}(14.5,7)

\psline[linestyle=dashed, dash=3pt 2pt]{->}(0.2,4)(6.5,4)

\rput(1.2,4.2){$a$}%
\psdots(1.5,4) \rput(5.7,3.8){$b$} \psdots(5.5,4)
\pscurve(1.5,4)(1.5,3.8)(1.7,3)(2.5,2.4)(3,1.8)(4,1.3)(4.3,1.7)(4.6,2)(5,3)(5.5,3.8)(5.5,4)
(5.5,4.2)(5.7,4.4)(5.2,4.7)(4.4,5.5)(4.2,5.7)(3.7,6.3)(3.3,6)(3,6.5)(2.5,6)(2.2,5)(1.7,4.3)(1.5,4.2)(1.5,4)

\psline[linewidth=2pt]{<-}(2.45,2.45)(2.55,2.35)
\rput(2.4,2.2){$\Gamma_{\da}$}
\psline[linewidth=2pt]{<-}(4.45,5.45)(4.35,5.55)
\rput(4.6,5.7){$\Gamma_{\ua}$}

\rput(0.7,6){$\Upsilon=\Xi$} \rput(3.5,5){$\Upsilon=\Xi M_{\ua}$}
\rput(3,3){$\Upsilon=\Xi M_{\da}^{-1}$}

\pscurve[linewidth=1pt]{->}(8.2,6)(7.3,6.5)(6.3,6)
\rput(7.3,6.2){$p^{-1}$}

\psline[linestyle=dashed, dash=3pt 2pt]{->}(8,4)(13.7,4)
\psdots(8.5,4)(13,4) \rput(7.7,3.7){$p(a)$}
\rput(13.5,3.7){$p(b)$}

\psline{-}(8.5,2.5)(8.5,5.5)
\pscurve{-}(8.5,5.5)(8.65,5.9)(9,6)

\psline{-}(9,6)(12.5,6)
\pscurve{-}(12.5,6)(12.85,5.9)(13,5.5)

\psline{-}(13,5.5)(13,2.5)
\pscurve{-}(13,2.5)(12.85,2.1)(12.5,2)

\psline{-}(12.5,2)(9,2)
\pscurve{-}(9,2)(8.65,2.1)(8.5,2.5)


\rput(11,5){$p(H_I)$}

\rput(11,3){$p(H_{II})$}
\rput(7.7,5){$p(H_{III})$}


\rput(3,4.4){$H_I$}

\rput(3.5,1.8){$H_{II}$}
\rput(5.5,5){$H_{III}$}

\psline[linewidth=2pt]{<-}(12,2)(12.1,2)
\psline[linewidth=2pt]{<-}(11,6)(10.9,6)

\end{pspicture}
\caption{Contours $\Gamma_{\ua}$ and
$\Gamma_{\da}$ associated with the RHP for $\Upsilon$. The second figure depicts how $p$ maps the contours
$\Ga_{\da}$ and $\Ga_{\ua}$. \label{contour pour le RHP de Y}}
\end{center}
\end{figure}

\section{The parametrices}
\label{Section parametrices}
\subsection{Parametrix around $a$}

The local parametrix $\mc{P}_a=\e{id} + \wh{\mc{P}}_a$ on a small disk
$\mc{D}_{a,\delta}\subset U$ of radius $\delta$ and centred at $a$,
 is an exact solution of the RHP:
\begin{itemize}
\item $\wh{\mc{P}_a}(\la)$ is a holomorphic in $\la \in\mc{D}_{a,\delta}\setminus \big\{ \Gamma_{\ua} \cup \Gamma_{\da} \big\} $
integral operator on $L^{2}\big( \R^+, \dd s \big) \oplus L^{2}\big( \R^+, \dd s \big) $
\item  $\wh{\mc{P}_a}(\la)$ admits continuous $\pm$-boundary values $\big(\wh{\mc{P}_a}\big)_{\pm} (\la)$  on 
$ \big\{ \Gamma_{\ua} \cup \Gamma_{\da} \setminus\{a\}  \big\} \cap \mc{D}_{a,\delta}$; 
\item $\wh{\mc{P}_a}(\la)$ has the same singular structure as $\Ups$ around $\la = a $;
\item uniformly in $(s,s^{\prime}) \in \R^+\times \R^+$ and $\la \in \partial \mc{D}_{a,\de} $, one has 
\beq
\big| \big| \wh{\mc{P}}_a\big( \la \mid s,s^{\prime}\big) \big| \big| \; \leq \; \f{ C }{ x^{1-\veps}  } \cdot \ex{ - \f{c}{4} (s+s^{\prime}) } \quad \e{for} \; \e{some} 
\quad C>0 \; ; 
\label{ecriture RHP param borne sur bord disque}
\enq
\item $\left\{ \ba{ll}
   \mc{P}_{a;+}(\la) \cdot M_{\ua}(\la) \; = \; \mc{P}_{a;-}(\la) \quad
   &\text{for }\la \in \Gamma_{\ua} \cap \mc{D}_{a,\delta}, \vspace{2mm} \\
   \mc{P}_{a;+}(\la) \cdot  M_{\da}^{-1}(\la)=\mc{P}_{a;-}(\la) \quad
   &\text{for }\la \in \Gamma_{\da} \cap \mc{D}_{a,\delta}.
                \ea \right.$
\end{itemize}
Here $\veps_a= 2 \!\underset{ \la \in \partial \mc{D}_{a,\delta}}{\sup}\!\!\abs{\Re \big(\nu(\la) \big)}< 1$. The canonically
oriented contour $\partial \mc{D}_{a,\delta}$ is depicted in
Fig.~\ref{contours for the RHP for P}.
\begin{figure}[h]
\begin{center}

\begin{pspicture}(8.2,5)
\pscurve{-}(3,0.8)(3.5,1)(4,2.35)(4,2.4)(4,2.5)(4,2.6)(4,2.65)(4.6,4.4)
\pscircle(4,2.5){2}
\psline[linewidth=2pt]{->}(6,2.5)(6,2.6)

\rput(4.3,2.5){$a$}
\psdots(4,2.5)

\rput(3.7,3.5){$\Gamma_{\ua}$}
\psline[linewidth=2pt]{->}(4.15,3.2)(4.2,3.32)

\rput(4.4,1.5){$\Gamma_{\da}$}
\psline[linewidth=2pt]{->}(3.89,1.7)(3.94,1.85)

\psline{->}(7.1,2.5)(8,2.5) \rput(8,2.2){$\Re (\la)$}

\psline{->}(7.1,2.5)(7.1,3.4) \rput(6.7,3.2){$\Im (\la)$}

\end{pspicture}

\caption{Contours in the RHP for $\mc{P}_a$.\label{contours for the RHP for P}}
\end{center}
\end{figure}

\noindent Let $\zeta_{a}(\la)=x\big( p(\la)-p(a) \big)$ with $\e{arg}[\zeta_{a}(\la)]\in \intoo{-\pi}{\pi}$ for $\la \in \mc{D}_{a,\de}\setminus\intof{a- \de}{a}$
and set 
\begin{equation}
\mc{P}_a(\la) \; = \; \Psi_a(\la) \cdot \big[ \zeta_{a}(\la) \big]^{-\nu(\la)\sg_3} \cdot \ex{ \f{ \i \pi\nu(\la) }{2} }\cdot L_a(\la)\; + \;  
\left( \ba{cc}   \e{id}-\bs{O}_{11}(\la) &  0  \\  
			  0    		& \e{id}-\bs{O}_{22}(\la) \ea \right) \; . 
\label{parametrice en -q1}
\end{equation}
Above, we agree upon 
\begin{equation}
\Psi_a(\la)= 
              \begin{pmatrix}
                        \Psi\big(-\nu(\la),1; \ex{-\i \f{\pi}{2} } \zeta_{a}(\la) \big)  \cdot \bs{O}_{11}(\la)  
				& \i b_{12}(\la) \cdot \Psi\big(1+\nu(\la),1; \ex{\i \f{\pi}{2} } \zeta_{a}(\la) \big) \cdot \bs{O}_{12}(\la) \\
  -\i b_{21}(\la)\cdot \Psi\big( 1-\nu(\la),1; \ex{-\i \f{\pi}{2} } \zeta_{a}(\la) \big) \cdot \bs{O}_{21}(\la) & \Psi\big(\nu(\la),1; \ex{\i \f{\pi}{2} } \zeta_{a}(\la) \big)  \cdot \bs{O}_{22}(\la) 
              \end{pmatrix} \; ,
\label{ecriture parametrice en a}
\end{equation}
with
\beqa
b_{12}(\la)&=& - \i \f{ \sin \big[ \pi \nu(\la) \big] \cdot \Gamma^{2}\big( 1 + \nu(\la) \big) }
				      { \pi \a^2_{0}(\la) \cdot\big[ \zeta_{a}(\la)\big]^{ 2\nu(\la) }\ex{-2 \i \pi \nu(\la)}  }   \cdot \ex{ ixp(a) }  \,  , \\
 b_{21}(\la) & = & -\i \f{ \pi \a^2_{0}(\la) \cdot  \big[ \zeta_{a}(\la)\big]^{ 2\nu(\la) }\ex{-2 \i \pi \nu(\la)} }
					  { \sin\big[ \pi \nu(\la) \big] \cdot  \Gamma^{2}\big( \nu(\la) \big)  }  \cdot \ex{-\i x p(a) }  \, .
\eeqa
In \eqref{ecriture parametrice en a}, $\Psi(a,c;z)$ denotes the Tricomi confluent hypergeometric function
(CHF) of the second kind (see Appendix~\ref{Apendix CHF})  with the convention of choosing the cut along $\R^-$. 
The function $\Psi(a,c;z)$ admits an analytical continuation on the universal covering of ${\Bbb C}\setminus \{0\}$ and satisfies 
there monodromy relations \eqref{cut-Psi-1} - \eqref{cut-Psi-2} together with the asymptotic property \eqref{asy-Psi}.
Also, we have introduced the new function $\a_0$  by the equations,
\beq
\a_{0}(\la) \; = \; \a(\la)
\left\{ \ba{ll}
  1 \quad
   &\text{for }\la \in \mc{D}_{a,\delta}, \quad \Im \la > 0\vspace{2mm} \\
 \ex{2 \i \pi \nu(\la)}    \quad
   &\text{for }\la \in  \mc{D}_{a,\delta}, \quad \Im \la < 0.
                \ea \right.
\enq
which is a holomorphic function on $\mc{D}_{a,\delta}\setminus \intoo{a-\de}{ a}$. 
Finally, the expression for the piecewise holomorphic constant matrix $L_a(\la)$ depends on the
region of the complex plane. Namely,
\begin{equation}
L_a(\la) \; = \;  \left\{\ba{cc}  I_2 \otimes \e{id}                           & -\tf{\pi}{2}<\e{arg}\big[p(\la)-p(a) \big] < \tf{\pi}{2} \vspace{4mm},\\
\left( \ba{cc} \e{id}  & - \ex{-2 \i \pi \nu(\la)}\bs{P}(\la)\ex{ixp(\la)}  \\
                                0& \e{id} 
                                    \ea \right)  & \tf{\pi}{2}<\e{arg}\big[ p(\la)-p(a) \big] < \pi \vspace{4mm},\\
\left( \ba{cc}  \e{id} &0 \\
                             - \bs{Q}(\la)\ex{-ixp(\la)}& \e{id}
                                    \ea \right)  & -\pi<\e{arg}\big[ p(\la) - p(a) \big] < -\tf{\pi}{2}.
        \ea \right.
\label{matrice L constante}
\end{equation}
Using \eqref{asy-Psi},  \eqref{cut-Psi-1} and \eqref{cut-Psi-2} together with the relations,
\beq
\bs{O}_{jl}(\la) \cdot \bs{O}_{lk}(\la) \, = \, \bs{O}_{jk}(\la)  \; . 
\enq
one checks that our choice of the matrix $L_a$  implies that  $\mc{P}_{a}$ has the desired form of its asymptotic behaviour on the boundary $\Dp{}\mc{D}_{a,\de}$ while the  desired jump conditions are satisfied automatically. Furthermore,
referring again to \eqref{asy-Psi},  \eqref{cut-Psi-1}, one can see that the function $\mc{P}_{a}$ is continuous across the cut
$\intoo{a-\de}{ a}$
 and thus indeed equation \eqref{parametrice en -q1} determines a parametrix around  
$\la = a$.
\subsection{Parametrix around $b$}

The RHP for the parametrix $\mc{P}_b=I_2\otimes \e{id} \; + \; \wh{\mc{P}}_b $ around $b$ reads

\begin{itemize}
\item $\wh{\mc{P}_b}(\la)$ is a holomorphic in $\la \in\mc{D}_{b,\de}\setminus \big\{ \Gamma_{\ua} \cup \Gamma_{\da} \big\} $ 
integral operator on $L^{2}\big( \R^+, \dd s \big) \oplus L^{2}\big( \R^+, \dd s \big)$;
\item $\wh{\mc{P}_b}(\la)$  admits $L^2$ $\pm$-boundary values $\big( \wh{\mc{P}_b} \big)_{\pm}(\la)$ on 
$ \big\{ \Gamma_{\ua} \cup \Gamma_{\da} \setminus\{b\}  \big\} \cap \mc{D}_{b,\delta}$; 
\item $\wh{\mc{P}_b}(\la)$ has the same singular structure as $\Ups$ around $\la =b$;
\item uniformly in $(s,s^{\prime}) \in \R^+\times \R^+$ and $\la \in \partial \mc{D}_{b,\de} $, one has 
\beq
\big| \big| \wh{\mc{P}}_b\big( \la \mid s,s^{\prime}\big) \big| \big| \; \leq \; \f{ C }{ x^{1-\veps_b }  } \cdot \ex{ - \f{c}{4} (s+s^{\prime}) } 
\quad \e{for} \; \e{some} \quad C>0  ;
\enq
\item $\left\{ \ba{ll}
   \mc{P}_{b;+}(\la) \cdot M_{\ua}(\la) \; = \; \mc{P}_{b;-}(\la) \quad
   &\text{for }\la \in \Gamma_{\ua} \cap \mc{D}_{b,\delta} \,  , \\
   \mc{P}_{b;+}(\la) \cdot M_{\da}^{-1}(\la)=\mc{P}_{b;-}(\la) \quad
   &\text{for }\la \in \Gamma_{\da} \cap \mc{D}_{b,\delta} \, 
                \ea \right.$ \;  ; 
\end{itemize}
and $\veps_b=2\sup_{\la \in \partial D_{b,\delta}}\abs{\Re(\nu(\la))}<1$ .
\begin{figure}[h]
\begin{center}

\begin{pspicture}(8.2,5)
\pscurve{-}(3,0.8)(3.5,1)(4,2.35)(4,2.4)(4,2.5)(4,2.6)(4,2.65)(4.6,4.4)
\pscircle(4,2.5){2}
\psline[linewidth=2pt]{->}(6,2.5)(6,2.6)

\rput(4.3,2.5){$b$}
\psdots(4,2.5)

\rput(3.7,3.5){$\Gamma_{\ua}$}
\psline[linewidth=2pt]{<-}(4.15,3.2)(4.2,3.32)

\rput(4.4,1.5){$\Gamma_{\da}$}
\psline[linewidth=2pt]{<-}(3.89,1.7)(3.94,1.85)

\psline{->}(7.1,2.5)(8,2.5) \rput(8,2.2){$\Re (\la)$}

\psline{->}(7.1,2.5)(7.1,3.4) \rput(6.7,3.2){$\Im (\la)$}

\end{pspicture}

\caption{Contours in the RHP for $\mc{P}_b$.\label{Contours for RHP for P tilde}}
\end{center}
\end{figure}

\noindent Note that the solution to the RHP for the parametrix $\mc{P}_b$ around $b$ can be formally obtained from the one at
$a$ through the transformation $b \rightarrow a$ and $\nu \tend -\nu$ on the solution to the RHP for $\mc{P}_a$.
 Indeed, the two RHP are identical modulo this negation. Just as for the parametrix around $a$, we focus on the  solution
\begin{equation}
\mc{P}_b(\la) \; =\;  \Psi_b(\la) \cdot \big[\zeta_{b}(\la) \big]^{\nu(\la)\sg_3}\ex{ - \f{ \i \pi\nu(\la) }{2} }  \cdot L_b(\la) \cdot \big[\zeta_{b}(\la) \big]^{\nu(\la)\sg_3}  \; + \;  
\left( \ba{cc}   \e{id}-\bs{O}_{11}(\la) &  0  \\  
			  0    		& \e{id}-\bs{O}_{22}(\la) \ea \right) 
\end{equation}
where $\zeta_b(\la)=x\big[p(\la)-p(b) \big]$ with $\e{arg}[\zeta_{b}(\la)]\in \intoo{-\pi}{\pi}$ for $\la \in \mc{D}_{b,\de}\setminus\intof{b- \de}{f}$, and 
\begin{equation}
\Psi(\la)= 
              \begin{pmatrix}
                        \Psi\big(\nu(\la),1; \ex{-\i \f{\pi}{2}} \zeta_{b}(\la) \big) \cdot  \bs{O}_{11}(\la)  & \i \tilde{b}_{12}(\la) \cdot 
				  \Psi\big(1-\nu(\la),1; \ex{\i \f{\pi}{2}} \zeta_{b}(\la) \big) \cdot \bs{O}_{12}(\la) \\
  -\i \tilde{b}_{21}(\la)\cdot \Psi\big( 1+\nu(\la),1; \ex{-\i \f{\pi}{2}} \zeta_{b}(\la) \big) \cdot \bs{O}_{21}(\la) & \Psi\big(-\nu(\la),1; \ex{-\i \f{\pi}{2}} \zeta_{b} (\la) \big) \cdot \bs{O}_{22}(\la) 
              \end{pmatrix}
%
,
\end{equation}
with
\beqa
\tilde{b}_{12}(\la)&=&  \i \f{ \sin \big[ \pi \nu(\la) \big] \Gamma^{2}\big( 1 - \nu(\la) \big) }
				      { \pi \a^2(\la)  } \cdot \big[ \zeta_b(\la) \big]^{ 2\nu(\la) }   \cdot \ex{ ixp(b) } \,  ,\\
 \tilde{b}_{21}(\la) &=& \i \f{ \pi \a^2(\la) \cdot \ex{-\i x p(b) }   }
				 { \sin\big[ \pi \nu(\la) \big]  \Gamma^{2}\big( -\nu(\la) \big)  \cdot \big[ \zeta_b(\la)  \big]^{2\nu(\la) }  }   \, .
\eeqa
Finally, the parametrix $L_b(\la)$ reads
\begin{equation}
L_b(\la) \; = \;  \left\{\ba{cc}  I_2 \otimes \e{id}                           & -\tf{\pi}{2}<\e{arg}\big[p(\la)-p(b) \big] < \tf{\pi}{2} \vspace{4mm},\\
\left( \ba{cc} \e{id}  & - \bs{P}(\la)\ex{ixp(\la)}  \\
                                0& \e{id} 
                                    \ea \right)  & \tf{\pi}{2}<\e{arg}\big[ p(\la)-p(b) \big] < \pi \vspace{4mm},\\
\left( \ba{cc}  \e{id} &0 \\
                             - \bs{Q}(\la)\ex{-ixp(\la)}& \e{id}
                                    \ea \right)  & -\pi<\e{arg}\big[ p(\la) - p(b) \big] < -\tf{\pi}{2}.
        \ea \right.
\label{matrice L constante2}
\end{equation}

\subsection{The last transformation}

We define the integral operator $\Pi(\la) = \e{id}\otimes I_2\, + \, \wh{\Pi}(\la)$ as 
\begin{equation}
\Pi(\la) \; = \; \left\{  \ba{ll}
                    \Upsilon(\la) \cdot \mc{P}^{-1}_{b}(\la) &\text{for } \la\in \mc{D}_{b,\delta} \, ,\\
                    \Upsilon(\la) \cdot \mc{P}^{-1}_{a}(\la) &\text{for } \la \in \mc{D}_{a,\delta} \, , \\
                    \Upsilon(\la) &\text{for } \la \in \mathbb{C}\setminus\big\{ \ov{\mc{D}}_{a,\delta}\cup \ov{\mc{D}}_{b,\delta} \big\} \, .
            \ea \right.
\end{equation}
\begin{figure}[h]
\begin{center}

\begin{pspicture}(10,6.5)

\pscircle(2,4){1}
\psdots(2,4)
\rput(1.6,4){$a$}
\psline[linewidth=2pt]{->}(3,4)(3,4.1)

\pscircle(8,4){1}
\psdots(8,4)
\rput(8.2,4){$b$}
\psline[linewidth=2pt]{->}(9,4)(9,4.1)

\psline[linearc=.25]{-}(2,5)(2,6)(8,6)(8,5)
\psline[linewidth=2pt]{->}(5,6)(5.1,6)
\rput(5.2,5.2){$\Gamma_{\ua}'$}

\psline[linearc=.25]{-}(2,3)(2,2)(8,2)(8,3)
\psline[linewidth=2pt]{<-}(4,2)(4.1,2)
\rput(4.5,2.5){$\Gamma_{\da}'$}



\rput(3.5,1){$\Sg_\Pi=\Ga_{\da}'\cup\Ga_{\ua}'\cup\Dp{}\mc{D}_{a,\de} \cup \Dp{}\mc{D}_{b,\de}$}
\end{pspicture}

\caption{Contour $\Sg_\Pi$ appearing in the RHP for $\Pi$.\label{Contour for RHP for R}}
\end{center}
\end{figure}

\noindent It is readily checked that $\Pi=I_2\otimes \e{id} \; + \; \wh{\Pi} $  satisfies the Riemann--Hilbert problem 

\begin{itemize}
\item $ \wh{\Pi}(\la)$ is a holomorphic in $\la \in\Cx \setminus \Sg_{\Pi} $ 
integral operator on $L^{2}\big( \R^+, \dd s \big) \oplus L^{2}\big( \R^+, \dd s \big)$;
\item $\wh{\Pi}(\la)$  admits continuous $\pm$-boundary values $\big( \wh{\Pi} \big)_{\pm}(\la)$ on 
$ \Sg_{\Pi} $; 
\item uniformly in $(s,s^{\prime}) \in \R^+\times \R^+$ and for any compact $K$ such that $\Sg_{\Pi}\subset \e{Int}(K)$,
one has 
\beq
\big| \big| \wh{\Pi}\big( \la \mid s,s^{\prime}\big) \big| \big| \; \leq \; \f{ C }{ x^{1-\veps }  } \cdot \ex{ - \f{c}{4} (s+s^{\prime}) } 
\quad \ba{l} \e{for} \; \e{some} \quad C>0  \; , \; \; \e{any} \; \la \in \Cx \setminus K \vspace{2mm} \\
\e{and} \; \e{for} \; \veps = \max\{\veps_a, \veps_b\} \ea \; ; 
\label{ecriture bornes uniformes sur noyau integral de Pi}
\enq
\item  $\Pi_+(\la) \cdot G_{\Pi}(\la) \, =  \,   \Pi_-(\la) $ \qquad  where \qquad  $ G_{\Pi}(\la) \, = \, \left\{ \ba{ll}
M_{\ua}(\la) \;\;  \text{for } \; \la \in \Gamma_{\ua}^{\prime}  \; ;  \vspace{1mm} \\ 
M_{\da}^{-1}(\la) \; \;  \text{for }  \; \la \in \Gamma_{\da}^{\prime} \;  ;     \vspace{1mm} \\
   \mc{P}_{\vsg} (\la) \; \e{for} \; \la \in \Dp{}\mc{D}_{ \vsg,\de}  \; \;\e{with} \;  \vsg \in \big\{ a,b \big\}  \; . 
                \ea \right.$  
\end{itemize}

\begin{prop}
 
 The solution to the Riemann--Hilbert problem for $\Pi$ exists and is unique, provided that $x$ is large enough
 and $|\Im(t)| <\de$, with $\de>0$ but small enough.

\end{prop}

\Proof

The unique solvability of the Riemann--Hilbert problem for $\Pi$ is established along the lines already discussed. 
We hence solely focus on the existence of solutions. Introduce the following operator  
\beq
\mc{C}_{\Sg_{\Pi} } \big[ M \big] (\la) \, = \, \Int{ \Sg_{\Pi} }{}  \f{ M(\mu)   }{ \mu - \la } \cdot \f{ \dd \mu }{ 2\i \pi}
\; \; \e{for} \; \la \, \in \, \Cx \setminus \Sg_{\Pi} \quad \e{and}\quad  M \in\mc{M}_2\Big(L^2(\Sg_{\Pi}\times \R^+\times \R^+)\Big) \;. 
\enq
Then, we consider the below singular integral equation  for the unknown matrix $\wh{\Pi}_+\in\mc{M}_2\Big(L^2(\Sg_{\Pi}\times \R^+\times \R^+)\Big) $:
\beq
\wh{\Pi}_+(\la) \; + \; \mc{C}_{\Sg_{\Pi};+}\big[ \wh{\Pi}_+ \wh{G}_{\Pi} \big] (\la) \; = \; - \mc{C}_{\Sg_{\Pi};+}\big[ \wh{G}_{\Pi} \big] (\la) 
\qquad \e{where} \quad G_{\Pi}(\la) \, = \, \e{id}\otimes I_2 \, + \, \wh{G}_{\Pi}(\la) \;. 
\enq
It follows from 
\beq
\wh{G}_{\Pi} \in \mc{M}_2\big(L^2\cap L^{\infty}(\Sg_{\Pi}\times \R^+\times \R^+)\big)  \quad \e{with} \quad 
\norm{  \wh{G}_{\Pi} }_{ \mc{M}_2\big(L^2\cap L^{\infty}(\Sg_{\Pi}\times \R^+\times \R^+)\big) } \; \leq \; 
 \f{ C }{ x^{1-\veps} } \;, 
\enq
that, for any $M\in \mc{M}_2\big(L^2\cap L^{\infty}(\Sg_{\Pi}\times \R^+\times \R^+)\big)$, one has 
$M \wh{G}_{\Pi}  \in \mc{M}_2\big(L^2(\Sg_{\Pi}\times \R^+\times \R^+)\big) $ Furthermore, 
 one has that $\la \mapsto \big(M \wh{G}_{\Pi} \big)(\la\mid s, s^{\prime})$
belongs to $\mc{M}_2\big(L^2(\Sg_{\Pi})\big)$ almost everywhere in $(s,s^{\prime}) \in \R^+\times \R^+$. 
Therefore, using Fubbini's theroem and the continuity of the $+$ boundary value of the Cauchy operator on $\Sg_{\Pi}$ in
respect to the $L^2(\Sg_{\Pi})$ norm, we get 
\bem
\big| \big| \mc{C}_{\Sg_{\Pi};+}\big[ M \wh{G}_{\Pi} \big] \big| \big|^2_{ \mc{M}_2\big(L^2(\Sg_{\Pi}\times \R^+\times \R^+)\big) }
\; = \; \Int{\R^+\times\R^+}{} \dd s \dd s^{\prime} \bigg\{  
\big| \big| \mc{C}_{\Sg_{\Pi};+}\big[ (M \wh{G}_{\Pi})(*\mid s,s^{\prime}) \big] \big| \big|^2_{ \mc{M}_2\big(L^2(\Sg_{\Pi})\big) }   \bigg\} \\
\; \leq \; c_{\Pi} \cdot \big| \big|  M \wh{G}_{\Pi}  \big| \big|^2_{ \mc{M}_2\big(L^2(\Sg_{\Pi}\times \R^+\times \R^+)\big) }
\; \leq \; \f{ C_{\Pi} }{x^{1-\veps} }\big| \big|  M \big| \big|^2_{ \mc{M}_2\big(L^2(\Sg_{\Pi}\times \R^+\times \R^+)\big) } \;. 
\end{multline}
This guarantees the invertibility of the operator $\e{id}\otimes I_2 +  \mc{C}_{\Sg_{\Pi};+}\big[ \cdot  \wh{G}_{\Pi} \big] $ on 
$\mc{M}_2\big(L^2(\Sg_{\Pi}\times \R^+\times \R^+)\big)$. 
Since $\mc{C}_{\Sg_{\Pi};+}\big[ \wh{G}_{\Pi} \big] \in \mc{M}_2\big(L^2(\Sg_{\Pi}\times \R^+\times \R^+)\big)$, it follows 
that $\wh{\Pi}_+$ exists and that, furthermore, 
\beq
\big| \big|  \wh{\Pi}_+  \big| \big|^2_{ \mc{M}_2\big(L^2(\Sg_{\Pi}\times \R^+\times \R^+)\big) } \; \leq \; \f{ C }{x^{1-\veps} }
\quad \e{for} \; \e{some} \quad C>0 \;. 
\enq
We then define 
\beq
\Pi(\la) \; = \; \e{id}\otimes I_2  \, - \,  \mc{C}_{\Sg_{\Pi} } \big[ \wh{G}_{\Pi} \big] (\la) 
\, - \,  \mc{C}_{\Sg_{\Pi} } \big[ \wh{\Pi}_+ \wh{G}_{\Pi} \big] (\la) 
\;. 
\label{ecriture form explicite Pi via Pi+}
\enq
It is then straightforward, by using the bounds on $\wh{\Pi}_+$ and $\wh{G}_{\Pi}$, to deduce that $\Pi(\la)$ as defined through \eqref{ecriture form explicite Pi via Pi+}
does satisfy the Riemann--Hilbert problem stated above, with the sole difference that it admits $L^2$ $\pm$-boundary values on 
$\Sg_{\Pi}$. However, $\wh{G}_{\Pi}$ being a holomorphic integral operator on $L^2(\R^+,\dd s) \oplus L^2(\R^+,\dd s)$
in some open neighbourhood of $\Sg_{ \Pi}$ it is readily seen that $\wh{\Pi}_{\pm}$ admits a holomorphic continuation to 
some open neighbourhood of $\Sg_{\Pi}$  located on its $\mp$-side. In particular, this ensures that $\wh{\Pi}$ does admit, in fact, 
continuous $\pm$ boundary values on $\Sg_{\Pi}$. \qed

\section{The asymptotic behaviour of the determinant}
\label{Section calcul asymptotique log det}

\subsection{A determinant identity}

\begin{lemme}
 
 The following holds
\beq
\Dp{t}\ln\det\big[ I+V_t \big]  \; = \; \Oint{\Ga\big( \intff{a}{b} \big) }{} \hspace{-2mm}
z \cdot \e{tr}\Big[\Dp{z}\chi(z) \cdot \sg_3 \cdot \bs{\mf{s}} \cdot  \chi^{-1}(z)  \Big] \cdot \f{\dd z}{2\pi}
\qquad \e{where} \quad \sg_3 \; = \; \left(\ba{cc} 1 & 0 \\ 0 & -1 \ea \right)
\enq
and $\bs{\mf{s}}$ is the operator of multiplication by $s$, \textit{viz}. $\big(\bs{\mf{s}}\cdot f \big)(s) \, = \, s f(s)$. 
 Note that $\e{tr}$ appearing above refers to the matrix and operator trace. 
 
\end{lemme}

Note that the trace used above is well defined due to \eqref{ecriture bornes chi a l'infini} and the fact that $\wh{\chi}\big(\la\mid s, s^{\prime}\big)$
is smooth in all its variables for $\la$ uniformly away from $\intff{a}{b}$.

\Proof

Starting from the identity
\begin{equation}
\Dp{t}\ln \det[I+V_t] \; = \; \Int{a}{b} \big[ \Dp{t}V_t \cdot (I-R_t) \big](\la,\la) \cdot \dd\la \; ,
\end{equation}
along with
\begin{equation}
\Dp{t}V_t(\la,\mu) \; = \; 
-\Oint{ \Ga\big( \intff{a}{b} \big) }{} \hspace{-2mm} \f{\dd z}{2\pi} \cdot  \f{z}{(z-\la)(z-\mu)} \cdot
\Big( \vec{\bs{E}}_{L}(\la) , \bs{\mf{s}}  \sg_3 \vec{\bs{E}}_R(\mu) \Big) \; ,
\end{equation}
as well as invoking the representation of the resolvent $R_t$ in terms  of $\vec{\bs{F}}_L$ and $\vec{\bs{F}}_R$, we get
\begin{multline}
\Dp{t}\ln \det\big[I+V_t\big]
\; = \;  -\Oint{ \Ga\big( \intff{a}{b} \big) }{}\hspace{-4mm} \f{\dd z}{2\pi}\, z
\Int{a}{b} \dd \la \,  \f{ \Big( \vec{\bs{E}}_{L}(\la) , \bs{\mf{s}} \sg_3 \vec{\bs{E}}_R(\la) \Big) }{ (z-\la)^2 }\\
 +\e{tr}\Bigg\{ \Oint{ \Ga\big( \intff{a}{b} \big) }{}\hspace{-4mm} \f{\dd z}{4\pi}\, z
\Int{a}{b} \hspace{-2mm} \dd \la \dd \mu \,  \vec{\bs{F}}_R(\la)\otimes \Big(\bs{E}_{L}(\la) \Big)^{\bs{T}}
\bigg\{ \f{1}{\la-z} \, - \, \f{1}{\la-\mu} \bigg\} \f{ \bs{\mf{s}}  \sg_3 }{ (z-\mu)^2 }\vec{\bs{E}}_R(\mu)\otimes \Big(\vec{\bs{F}}_{L}(\mu)\Big)^{\bs{T}} \Bigg\}\;.
\end{multline}
By using the integral representation for $\chi$ , we obtain
\bem
\Dp{t}\ln \det\big[I+V_t\big]
\; = \;  -\Oint{ \Ga\big( \intff{a}{b} \big) }{}\hspace{-4mm} \f{\dd z}{2\pi}\, z
\Int{a}{b} \dd \la \, \f{ \Big( \vec{\bs{E}}_{L}(\la) , \bs{\mf{s}} \sg_3 \vec{\bs{E}}_R(\la) \Big) }{ (z-\la)^2 }\\
 + \Oint{ \Ga\big( \intff{a}{b} \big) }{}\hspace{-4mm} \f{\dd z}{2\pi}\, z
\Int{a}{b} \hspace{-2mm}  \dd \mu \,  \e{tr}\bigg\{ \Big[ \chi(\mu ) \, -\,  \chi(z )  \Big]
\cdot \f{ \bs{\mf{s}}  \sg_3 }{ (z-\mu)^2 }\vec{\bs{E}}_R(\mu)\otimes \Big(\vec{\bs{F}}_{L}(\mu )\Big)^{\bs{T}} \bigg\}\;.
\end{multline}
Finally, recalling the integral representation for $\chi^{-1}(\la)$, one gets 
\beq
\Dp{t}\ln \det\big[I+V_t\big] \; = \;  - \Oint{ \Ga\big( \intff{a}{b} \big) }{}\hspace{-2mm} 
 z \cdot \e{tr} \bigg\{ \chi(z) \cdot \sg_3 \bs{\mf{s}} \cdot \Dp{z}\chi^{-1}(z) \bigg\} \cdot \f{\dd z}{2\pi} \;. 
\enq
It solely remains to invoke that $\Dp{z}\Big(\chi^{-1}(z) \Big) \, = \,  - \chi^{-1}(z)  \cdot \Dp{z}\chi(z) \cdot \chi^{-1}(z)$
and the cyclic property of the trace. \qed

\subsection{The asymptotic evaluation of the determinant}

\begin{prop}
 
 The following representation holds for the ratio of determinants
\beq
\f{ \det\big[ I+ V_1 \big] }{\det\big[ I+ V_0\big]} \; = \;    \det\big[ I+ \mc{U}_{1;t=1} \big]\cdot \det\big[ I+ \mc{U}_{2;t=1} \big] 
\cdot \Big( 1 \,  + \, \e{o}(1)  \Big) \;. 
\enq

\end{prop}

\Proof 

Let $t$ be such that 
\beq
\det_{\Ga(\intff{a}{b})} \big[ \e{id}+\mc{U}_{k;t} \big] \; \not= \; 0 \qquad \e{for} \; k=1,2\;. 
\enq
Then, the Riemann--Hilbert analysis ensures that, uniformly away from $\intff{a}{b}$, the solution $\chi$ can be represented as 
\beq
\chi(\la) \; = \; \Big( I_2 \otimes \e{id} \; + \; \wh{\Pi} \Big) \cdot
	\left( \ba{cc } \be_1(\la) & 0 \\
				0   &  \be_2(\la) \ea \right) \;. 
\enq
where $\wh{\Pi}$ is an integral operator on $L^2\big( \R^+,\dd s  \big) \oplus L^2\big( \R^+,\dd s  \big)$ that, furthermore, satisfies to the bounds
\beq
 \Big( \wh{ \Pi }_{ak} \cdot \be_k \Big)\big( \la \mid s,s^{\prime} \big) \; \leq \; \f{  C \ex{-\f{c}{4}(s+s^{\prime})}  }{  x^{1-\veps} (1+|\la|) }
 \qquad \e{with}\quad  \veps \, =\, \max \{\veps_a, \veps_b \}
\enq
for $\la \in \Cx \setminus K$, with $K$ a small compact such that $\e{Int}(K) \supset \Sg_{\Pi}$, and any $s,s^{\prime} \in \R^+$.

As a consequence, one gets that 
\beq
\Dp{t} \ln \det\big[I+V_t] \; = \; \Oint{ \Ga( \Sg_{\Pi} ) }{}  \e{tr}\Big[ \Dp{z}\be_{1}(z) \cdot \bs{\mf{s}} \cdot \be^{-1}_{1}(z) \, - \, 
\Dp{z}\be_{2}(z) \cdot \bs{\mf{s}} \cdot \be^{-1}_{2}(z)   \Big]  \cdot z \f{ \dd z }{ 2\pi }  \; + \; \e{O}\Big( \f{1}{ x^{1-\veps} } \Big)
\enq
where the remainder $\e{O}(x^{\veps-1} )$ is in respect to the $x \tend +\infty$ limit.
Thus, by using the representation 
\beq
 \be_k(\la) \; = \; \e{id} \, - \,   \mc{C}\Big[  \tau_k \bs{\rho}_k \otimes \bs{\kappa}_k \Big](\la) 
 \qquad \e{where} \quad \bs{\rho_k(\mu)}(s) \; = \; \rho_k(\mu;s)
\enq
we are led to:
\beq
\Dp{t} \ln \det\big[I+V_t] \; = \; \sul{k=1}{2} \eps_k  
\Int{a}{b} \f{ \dd \mu }{2\pi}   \tau_k(\mu) \cdot \bs{\kappa}_k(\mu)\big[ \bs{\mf{s}}\bs{\rho}_k(\mu)\big]  \; + \;\e{O}(x^{\veps-1} )
\label{ecriture rep int pour Dt det}
\enq
where we remind that the function $\rho_k(\mu;s)$ is defined by 
\beq
\rho_k(\mu;s) \; = \; \Big( I+ \mc{K}_{k;t}\Big)^{-1}\big[ w_k(*;s) \big](\mu) \qquad \e{where} \qquad
w_k(\la;s) \; = \; \a_{k;+}(\la)   
\Oint{ \Ga(\intff{a}{b}) }{} \f{ \sqrt{c} \cdot \ex{-\f{c}{2}s-\i t \eps_k \mu s } }{ \a_k(\mu) \cdot (\mu-\la) } \cdot \f{ \dd \mu }{ 2\i \pi } \;. 
\enq
Note that, above, the operator $\Big( I + \mc{K}_{k;t} \Big)^{-1}$ acts on the $*$ variable of its argument. 
As a consequence,
\beq
\Dp{t} \ln \det\big[I+V_t] \; = \; - 
\sul{k=1}{2} \eps_k \Int{a}{b} \! \tau_k(\mu) \cdot \Big( I+ \mc{K}_{k;t} \Big)^{-1}\Big[ \bs{\kappa}_k(\mu)\big[ \bs{\mf{s}}w_k(*;\bullet)\big]  \Big](\mu) 
\cdot  \f{ \dd \mu }{2  \pi}  \; + \;\e{O}(x^{\veps-1} )\;. 
\enq
where the $*$ indicates the variable on which $\Big( I+\mc{K}_{k;t}\Big)^{-1}$ acts whereas the $\bullet$ variable refers to the one on which 
the one-form $\bs{\kappa}_k(\mu)$ acts. Observe that 
\bem
 \eps_k \tau_k(\mu)  \bs{\kappa}_k(\mu)\big[ \bs{\mf{s}}w_k(\nu;\bullet)\big] \; = \; c \eps_k \tau_k(\mu)  \a_{k;+}(\nu)   
 \Oint{ \Ga(\intff{a}{b}) }{} \hspace{-3mm} \f{ \dd \la }{ 2\i \pi } \f{ \a_k^{-1}(\la) }{ \la-\nu }  \Int{0}{+\infty}  \dd s  s \ex{- c s + \i t \eps_k (\mu-\la) s } \\ 
 \; = \;c \eps_k \tau_k(\mu)  \a_{k;+}(\nu)   \f{ \Dp{} }{ \Dp{} t  }
\bigg\{  \Oint{ \Ga(\intff{a}{b}) }{} \hspace{-3mm} \f{ \dd \la }{ 2\i^2 \pi } \f{ \a_k^{-1}(\la) }{ (\la-\nu)\eps_k (\mu-\la) } 
 \Int{0}{+\infty}  \dd s  \ex{- c s + \i t \eps_k (\mu-\la) s }   \bigg\}  \\
 \; = \; c \eps_k \tau_k(\mu)  \a_{k;+}(\nu)   \f{ \Dp{} }{ \Dp{} t  }   \bigg\{ \Oint{ \Ga(\intff{a}{b}) }{}
\f{ \a_k^{-1}(\la) }{ (\la-\nu) (\mu-\la) \big( t (\mu - \la) + \i \eps_k c \big) } \cdot  \f{ \dd \la }{ 2\i \pi }   \bigg\} \\ 
\; = \;  -\f{ \Dp{} }{ \Dp{} t  } \bigg\{   \f{ \a_{k;+}(\nu) }{ \a_{k;+}(\mu) }\cdot \f{ \a_{k;-}(\mu)-\a_{k;+}(\mu) }{  \a_k(\mu + \i \eps_k \tf{c}{t} ) } \cdot 
\f{   t  }{  \i  \big( t(\mu-\nu) \,+\, \i \eps_k c\big)  }   \bigg\}  \; = \; - 2\pi \cdot \Dp{t} \Big( K_{k;t}(\nu,\mu) \Big) \;. 
\end{multline}
Therefore, we get that 
\bem
\Dp{t} \ln \det\big[I+V_t] \; = \; - \sul{k=1}{2} \Int{a}{b} \Big( \big(I+\mc{K}_{k;t}\big)^{-1} \cdot \Dp{t} \mc{K}_{k;t} \Big)(\mu,\mu) \cdot \dd \mu 
\; + \; \e{O}\big( x^{\veps-1} \big)   \\
\; = \; -\f{\Dp{}}{ \Dp{}t} \ln \Big\{  \det\big[I+ \mc{K}_{1;t} \big] \cdot  \det\big[I+ \mc{K}_{2;t}  \big]  \Big\}  \; + \; \e{O}\big( x^{\veps-1} \big) \\ 
\; = \; 
-\f{\Dp{}}{ \Dp{}t} \ln \Big\{  \det_{\Ga(\intff{a}{b})} \big[I+ \mc{U}_{1;t} \big] \cdot  \det_{\Ga(\intff{a}{b})} \big[I+ \mc{U}_{2;t}  \big]  \Big\} 
\; + \; \e{O}\big( x^{\veps-1} \big)  \;. 
\label{ecriture formule asymptotique pour derivee log du det}
\end{multline}
Now, observe that there exists $\de>0$ such that 
\beq
t \mapsto \det_{\Ga(\intff{a}{b})}\big[I+ \mc{U}_{k;t} \big] \quad k=1,2 \; , 
\enq
are holomorphic functions on $\{ t\in \Cx \; : \; |\Re(t)|<2 \; \e{and}  \; |\Im(t)|<\de \}$ that furthermore
do not vanish at $t=0$ and $t=1$. 
As a consequence, it has a finite amount of zeroes located in $\{ t\in \Cx \; : \; |\Re(t)|< 1.5 \; \e{and}  \; |\Im(t)|< \tf{\de}{2} \}$. 
Thus, there exists a smooth curve $\msc{C}$ joining $0$ to $1$, located  in the region $|\Im(t)|< \tf{\de}{2}$ and such that 
\beq
 \det_{\Ga(\intff{a}{b})}\big[I+ \mc{U}_{k;t} \big] \not= 0 \quad \e{for} \; \e{any} \; t \in \msc{C} \; \e{and}\;\;  k=1,2 \; . 
\enq
As a consequence, the formula \eqref{ecriture formule asymptotique pour derivee log du det} holds for any $t \in \msc{C}$. 
Thence, integrating  both sides of \eqref{ecriture formule asymptotique pour derivee log du det} along $\msc{C}$
leads to the claim upon taking the exponent. Note that different choices of the curve $\msc{C}$ could lead to different values
of the integral. However, any two such integrals will differ by integer multiples of $2\i \pi$, hence leading to the same value of the 
exponents. \qed

\section*{Acknowledgements}

K. K. Kozlowski is supported by CNRS. He also acknowledges a funding from the ANR grant "DIADEMS" and
from the FABER-PARI burgundy region grant  "Asymptotique d'int\'{e}grales multiples". 
The work of A. R. Its was supported in
part by NSF grant DMS-1001777 and the  SPbGU grant   № 11.38.215.2014.
A. R. Its and  K. K. Kozlowski would like to thank the mathematics department of Technische Universit\"{a}t Berlin for its warm hospitality 
during the period where this work has been initiated.
K.K. Kozlowski is also indebted to Katedra Metod Matematyczny dla Fizyki of the Warsaw University, for providing 
excellent working conditions that allowed for finalising this work.


\appendix

\section{Some properties of confluent hypergeometric function}
\label{Apendix CHF}

For generic parameters $\pa{a,c}$ the Tricomi confluent hypergeometric
function $\Psi\pa{a,c;z}$ is one of the  solutions to the
differential equation
\begin{equation}
z y'' + \pa{c-z} y' -a y=0\;.
\end{equation}
It enjoys the monodromy properties 
\begin{align}
 &\Psi(a,1;z \ex{2 \i\pi} ) \;  = \;   \Psi(a,1;z) \ex{-2\i\pi a} \, + \, 
 \frac{2\pi \i \ex{-\i\pi a+z} }{\Gamma^2(a)}
\Psi(1-a,1; \ex{\i \pi} z)\, ,  
\label{cut-Psi-1}\\
 &\Psi(a,1;ze^{-2i\pi})= \Psi(a,1;z)e^{2i\pi a} \, - \, 
 \frac{ 2\pi \i \ex{\i \pi a+z} }{\Gamma^2(a)} \Psi(1-a,1;\ex{ - \i \pi } z) \, ,
\label{cut-Psi-2}
\end{align}
and has the asymptotic expansion:
\begin{equation}
 \Psi(a,c;z) \sim \sum_{n=0}^\infty(-1)^n\frac{(a)_n(a-c+1)_n}{n!}z^{-a-n},
 \quad z\to\infty,
 \quad -\frac{3\pi}2<\arg(z)<\frac{3\pi}2,
\label{asy-Psi}
\end{equation}
with $(a)_n \, = \,  \tf{ \Ga(a+n) }{ \Ga(a) }$.

\end{document}